\documentclass[aps,12pt,showpacs,amsfonts,amssymb,amsmath,notitlepage,tightenlines]{revtex4-1}
\usepackage{graphicx}																								
\usepackage{mathtools}																							
\usepackage{bm}																													
\usepackage{siunitx}																									
\usepackage{blindtext}																								
\usepackage{booktabs}																								
\usepackage{multirow}																							
\usepackage{color}																											

\graphicspath{{figures/}}

\begin{document}

\title{Linear and nonlinear instability in vertical counter-current laminar gas-liquid flows}

\author{Patrick Schmidt}
\affiliation{School of Engineering, The University of Edinburgh, Edinburgh, EH9 3JL, United Kingdom}

\author{Lennon \surname{\'{O} N\'{a}raigh}}
\affiliation{School of Mathematical Sciences, University College Dublin, Belfield, Dublin~4, Ireland}
\affiliation{Complex and Adaptive Systems Laboratory, University College Dublin, Belfield, Dublin~4, Ireland}

\author{Mathieu Lucquiaud}
\affiliation{School of Engineering, The University of Edinburgh, Edinburgh, EH9 3JL, United Kingdom}

\author{Prashant Valluri}
\email[Author to whom correspondence should be addressed. Electronic mail: ]{prashant.valluri@ed.ac.uk}
\affiliation{School of Engineering, The University of Edinburgh, Edinburgh, EH9 3JL, United Kingdom}

\date{\today}

\begin{abstract}
We consider the genesis and dynamics of interfacial instability in gas-liquid flows, using as a model the two-dimensional channel flow of a thin falling film sheared by counter-current gas. The methodology is linear stability theory (Orr-Sommerfeld analysis) together with direct numerical simulation of the two-phase flow in the case of nonlinear disturbances.
We investigate the influence of three main flow parameters (density contrast between liquid and gas, film thickness, pressure drop applied to drive the gas stream) on the interfacial dynamics. Energy budget analyses based on the Orr-Sommerfeld theory reveal various coexisting unstable modes (interfacial, shear, internal) in the case of high density contrasts, which results in mode coalescence and mode competition, but only one dynamically relevant unstable internal mode for low density contrast. The same linear stability approach provides a quantitative prediction for the onset of (partial) liquid flow reversal in terms of the gas and liquid flow rates. A study of absolute and convective instability for low density contrast shows that the system is absolutely unstable for all but two narrow regions of the investigated parameter space. Direct numerical simulations of the same system (low density contrast) show that linear theory holds up remarkably well upon the onset of large-amplitude waves as well as the existence of weakly nonlinear waves. In comparison, for high density contrasts corresponding more closely to an air-water-type system, although the linear stability theory is successful at determining the most-dominant features in the interfacial wave dynamics at early-to-intermediate times, the short waves selected by the linear theory undergo secondary instability and the wave train is no longer regular but rather exhibits chaotic dynamics and eventually, wave overturning.
\end{abstract}



\maketitle 

\section{Introduction}
\label{sec:introduction}
Vertical two-phase flows of thin liquid films sheared by a counter-current gas are prototypical for many technical applications, such as absorption and distillation (using structured packing), evaporation and condensation. In these applications, not only mass and heat transfer but also operational limits are closely linked to the prevailing hydrodynamics. The flow in the two phases, in turn, is largely determined by the interactions between gas and liquid at its interface. Although gas-sheared liquid films have been part of active research on fundamental and practical level for several decades, the rich interfacial dynamics are still not fully understood. These rich dynamics emerge as the liquid interface becomes unstable, leading to the development of waves. Depending on the flow rates, these waves can form highly complex structures, which may give rise to wave break-up, ligament formation and droplet entrainment.

One of the first to investigate a channel flow with two superimposed fluid layers from a theoretical point was Yih~\cite{Yih_1967}, who used asymptotic expansion to solve the Orr-Sommerfeld eigenvalue problem associated with the temporal linear stability in the long wavelength limit for equal densities and layer thicknesses. He found that viscosity stratification alone can cause interfacial instability at arbitrarily small Reynolds numbers, which is also referred to as Yih mechanism. Yiantsios~\cite{Yiantsios_1988} extended the linear stability analysis by accounting for short waves as well as effects due to surface tension and gravity. They observed that, depending on the choice of parameters, the flow is receptive to a short-wave instability at low Reynolds numbers and, moreover, to a shear mode instability (Tollmien-Schlichting mechanism) for sufficiently large Reynolds numbers. To classify the various types of instabilities arising in parallel two-phase flow, Boomkamp \textit{et al.}~\cite{Boomkamp_1996} analysed by which mechanism energy is transferred from the primary flow to growing disturbances, thereby verifying that both the Yih and shear mode are important routes to interfacial instability. In fact, a combination of these two mechanisms, also referred to as internal mode, represents a further possible route. An established method to obtain the full spectrum of eigenvalues in the Orr-Sommerfeld problem is the Chebyshev collocation method~\cite{Boomkamp_1997}. This method can also be used to determine whether the interface is convectively or absolutely unstable~\cite{Valluri_2010, ONaraigh_2013a}, a characteristic that is of importance for the operation of processes utilizing shear flows. An extensive review on the analysis of spatially developing flows can be found in Reference~\cite{Huerre_1990}.

Linear stability analysis is an effective and proven technique to gain valuable insight in the genesis of interfacial instability. However, its validity is limited to disturbances with infinitesimal amplitude. As these disturbances grow, nonlinear effects become important and have to be taken into account. Due to the complexity of nonlinear stability only few general theories exist~\cite{Schmid_2001}. Nonetheless, a variety of modelling techniques have been proposed to describe the development of the interfacial waves up to finite amplitude. Some of these techniques impose \textsl{a priori} assumptions about the wave dynamics under investigation, like long-wave or lubrication approximation, resulting in model equations such as the Kuramoto-Sivashinsky equation or depth-averaged integral equations. A large number of studies on the nonlinear dynamics of interfacial flows are based on these kind of models~\cite{Hooper_1985,Charru_1994,Tilley_1994,Dietze_2013}. However, given that their range of applicability is generally not known in advance, they may at times be prone to ambiguous results~\cite{Brevdo_1999}. This is especially the case for fairly ``punishing'' flow regimes, involving large pressure fluctuations and potentially large-amplitude waves, for which there is a major necessity to gain fundamental understanding. On the other hand, weakly nonlinear theories based on either the Stuart-Landau or the Ginzburg-Landau equations dispense with assumptions that cannot be confirmed a priori and are capable of matching experimental observations~\cite{Barthelet_1995, Sangalli_1997,Boomkamp_1998}. Such ``pure'' weakly nonlinear theories are therefore appropriate to complement direct numerical simulations (DNS) of the full Navier-Stokes equations, which, in turn, are guided by basic Orr-Sommerfeld (OS) analysis, to study interfacial instability in a rigorous manner.

The aforementioned techniques have helped to shed light on the genesis and development of interfacial instability in shear flows and a good understanding of the mechanisms at play has been gained. However, a substantial amount of the available literature is dedicated to horizontal flows or flows down an inclined plane. Flow dynamics specific to a vertical configuration have not received as much attention even though the same methods are applicable. Phenomena related to this configuration, such as partial liquid flow reversal due to counter-current gas (flooding), influence the design, optimization as well as operation of technical processes to a great extent and have hitherto mainly been investigated experimentally. We therefore try to help elucidate the rich dynamics of vertical counter-current gas-liquid flows from a more theoretical standpoint using the (semi-)analytical methods and simulation techniques described above.

In this work we consider the laminar-laminar case only, although the ideas and results contained herein can be extended to turbulent gas streams. The most accurate methodology appropriate for such an enhanced level of complexity is full-scale DNS, which is the target of future work. However, short of full-scale DNS, a quasi-laminar model may be assumed for the linear stability of the two-phase flow~\cite{ONaraigh_2011, ONaraigh_2013a} or, alternatively, a weighted residual integral boundary layer method (WRIBL) model~\cite{Ruyer-Quil_2000, Tseluiko_2011,Vellingiri_2015}, which also models nonlinear interfacial waves. These approaches both have their own advantages and shortcomings, and the aim of future work will be to confirm these reduced-dimensional modelling approaches with evidence from accurate DNS, towards which the present work is an initial contribution.

This work is organized as follows. We give a description of the investigated problem and outline the employed methods in \S~\ref{sec:problem_amp_methods}. Results regarding temporal stability of the system are discussed in \S~\ref{sec:temporal_stability}. Spatio-temporal behaviour with respect to absolute/convective instability of the linearized dynamics is presented in \S~\ref{sec:A/C_instability}. Section~\ref{sec:nonlinear_wave_dynamics} discusses nonlinear wave dynamics. Finally, we give concluding remarks in \S~\ref{sec:conclusions}.

\section{Problem description and computational methods}
\label{sec:problem_amp_methods}
\begin{figure}[htbp]
	\centering
		\includegraphics[scale=.7]{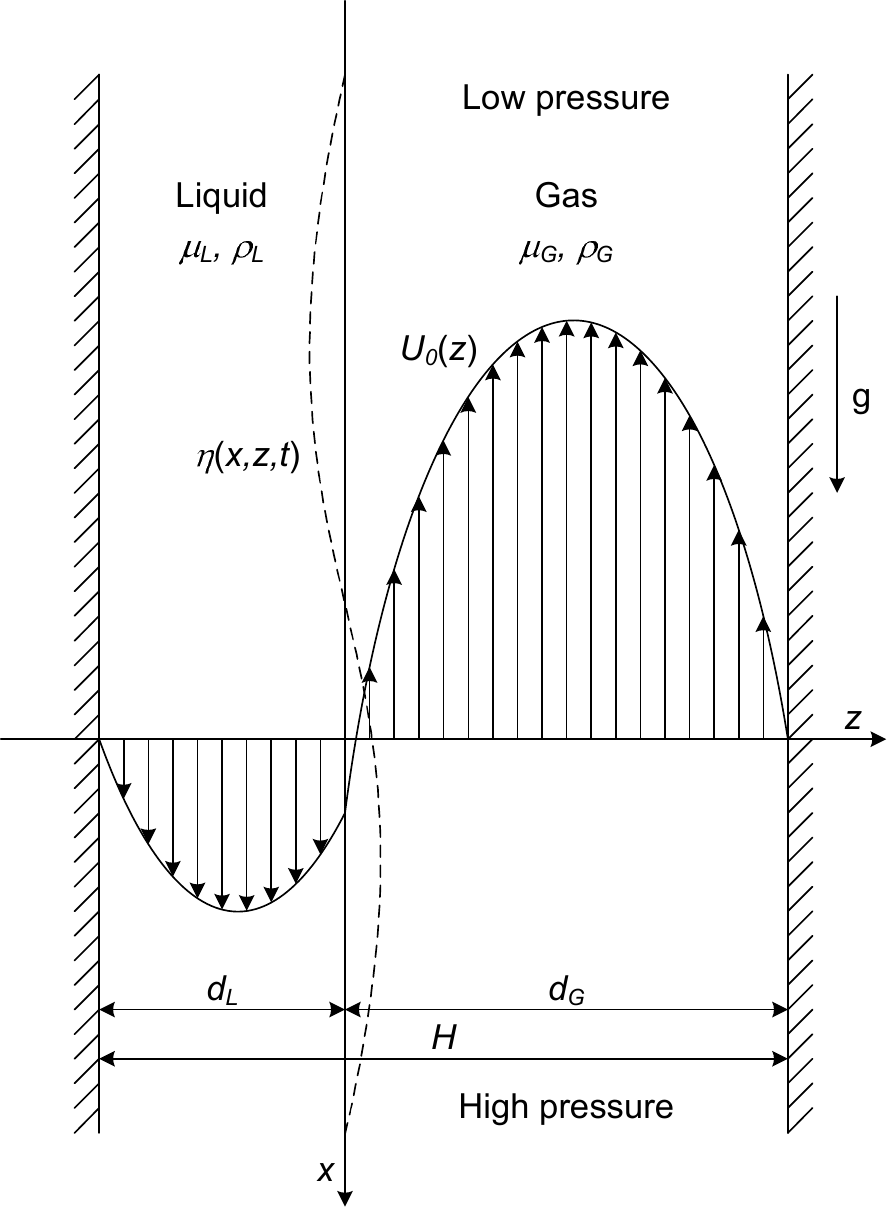}
	\caption{Schematic representation of the undisturbed base flow. Both fluids are assumed laminar. The dashed line indicates the development of a wave at the interface.}
	\label{fig:schematic_counter_laminar_laminar}
\end{figure}

In this analysis we consider the dynamics of a gas-liquid flow in a vertical channel, described schematically in Fig.~\ref{fig:schematic_counter_laminar_laminar}.
The two continuous phases are separated by an, initially, flat interface. A pressure gradient $\mathrm{d}p/\mathrm{d}x>0$ in vertical direction counteracts gravity. We investigate cases in which the balance between gravity and pressure gradient gives rise to counter-current flow, with gas flowing in the direction of decreasing pressure and liquid flowing in the direction of gravity. Figure~\ref{fig:schematic_counter_laminar_laminar} also shows the development of a linear small-amplitude wave at the interface. Typically, the evolution of interfacial waves depends sensitively on the details of the mean flow.

Both fluid layers exhibit steady, spatially uniform, laminar and incompressible flow along the vertical rectangular channel. To describe this two-dimensional flow, we use a Cartesian coordinate system, $\left(x,z\right)$, in which the flat interface is located at $z=0$ and the confining channel walls are located at $z = -d_{L}$ and at $z = d_{G}$, respectively. Within these boundaries, the fully developed liquid and gas layer occupy the regions $-d_{L}\leq z\leq 0$ and $0\leq z\leq d_{G}$, respectively.

\subsection{Base flow and linear stability analysis}
\label{subsec:base_flow_LSA}
With the above mentioned conditions, the Navier-Stokes equations describing the fluid flow in both phases reduce to standard balances between pressure, viscous and gravitational forces, which are subject to no-slip condition at the channel walls, $z = -d_{L}$ and $z = d_{G}$, as well as continuity of tangential stress at the interface, $z = 0$. To write the governing equations in nondimensional form we introduce the following dimensionless variables (without tildes) and scalings:
\begin{equation}
	\bm{\tilde{x}} = H\bm{x},\quad
	\bm{\tilde{u}} = V_{p}\bm{u},\quad
	\rho_{G}V_{p}^{2} = H\frac{\mathrm{d}p}{\mathrm{d}x},\quad
	\tau_{int} = \rho_{G}V_{\ast,int}^{2},\quad
	\tilde{t} = \frac{H}{V_{p}}t,
\label{eq:dimensionless_variables}
\end{equation}
where $\bm{x} = \left(x,z\right)$ and $\bm{u} = \left(u,w\right)$ are the coordinate and velocity vector, $H$ is the channel height, $V_{p}$ is an inertial pressure scale, $\mathrm{d}p/\mathrm{d}x$ is a positive pressure gradient, $\tau_{int}$ is the interfacial shear stress, $V_{\ast,int}$ is the gas side interfacial friction velocity and $t$ denotes time. Further, the following dimensionless parameters arise:
\begin{equation}
	\begin{gathered}
		m = \frac{\mu_{L}}{\mu_{G}},\enspace
		r = \frac{\rho_{L}}{\rho_{G}},\enspace
		\delta_{j} = \frac{d_{j}}{H},
		\\
		Re_{p} = \frac{\rho_{G}V_{p}H}{\mu_{G}},\enspace
		Re_{g} = \frac{\rho_{G}\sqrt{gH}H}{\mu_{G}},\enspace
		Re_{\tau} = \frac{\rho_{G}V_{\ast,int}H}{\mu_{G}},\enspace
		We = \frac{\rho_{G}V_{p}^{2}H}{\gamma}.
	\end{gathered}
\label{eq:dimensionless_parameters}
\end{equation}
Here, $\mu_{j}$ is the dynamic viscosity and $\rho_{j}$ the density of the respective phase $\left(j=L,G\right)$, whereas $\delta_{j}$ is the relative thickness of the respective fluid layers. The Reynolds numbers $Re_{p}$, $Re_{g}$ and $Re_{\tau}$, in turn, relate to the applied pressure drop, to gravity and to interfacial shear. A Weber number $We$ accounts for surface tension. With this rescaling, the velocity profile for the undisturbed base flow in nondimensional form reads
\begin{equation}
	u_{0}\left(z\right) =
	\left\{
	\begin{aligned}
		& \frac{1}{m}\left[\frac{1}{2}Re_{p}\left(1-r\frac{Re_{g}^{2}}{Re_{p}^{2}}\right)\left(z^{2}-\delta_{L}^{2}\right)
						- \frac{Re_{\tau}^{2}}{Re_{p}}\left(z+\delta_{L}\right)\right],&\quad-\delta_{L}\leq z\leq 0,	\\
		& \frac{1}{m}\left[-\frac{1}{2}Re_{p}\left(1-r\frac{Re_{g}^{2}}{Re_{p}^{2}}\right)\delta_{L}^{2}
						- \frac{Re_{\tau}^{2}}{Re_{p}}\delta_{L}\right]	\\
		& \phantom{\frac{1}{m}\left[\frac{1}{2}Re_{p}\left(1-r\frac{Re_{g}^{2}}{Re_{p}^{2}}\right)\left(z^{2}-\delta_{L}^{2}\right)
												- \frac{Re_{\tau}^{2}}{Re_{p}}\left(z+\delta_{L}\right)\right],}
				\mathllap{\left.+ \frac{1}{2}Re_{p}\left(1-\frac{Re_{g}^{2}}{Re_{p}^{2}}\right)z^{2}-\frac{Re_{\tau}^{2}}{Re_{p}}z\right.,}&\quad0 \leq z\leq \delta_{G}.
	\end{aligned}
\right.
\label{eq:velocity_profile}
\end{equation}
The ratio of pressure and gravity Reynolds number $Re_{p}/Re_{g}$ results in a Froude number, which, in combination with the pressure scale in Eq.~\eqref{eq:dimensionless_variables}, represents a measure for the effect of applied pressure drop relative to gravity acting on the gas layer:
\begin{equation}
	Fr = \frac{V_{p}}{\sqrt{gH}}
			= \sqrt{\frac{\frac{\mathrm{d}p}{\mathrm{d}x}}{\rho_{G}g}}.
\label{eq:froude_number}
\end{equation}

To gain insight into the development of small disturbances $\eta(\bm{x},t)$ centred around the flat interface $z = 0$, we examine the linear stability of the interface in terms of a standard Orr-Sommerfeld-type analysis (a comprehensive formulation of this analysis can be found in Appendix~\S\ref{app:perturbation_equations}. In this approach, the governing equations are linearized around the base flow, Eq.~\eqref{eq:velocity_profile}, and infinitesimally small perturbations in the associated streamfunction $\phi\left(z\right)$ are assumed to have wave-like solutions of the form $\Phi(x,z,t) = \mathrm{e}^{\mathrm{i}(\alpha x-\omega t)}\phi(z)$, where $\alpha = \alpha_{r} + \mathrm{i}\alpha_{i}$ is the complex wavenumber and $\omega = \omega_{r} + \mathrm{i}\omega_{i}$ is the complex angular frequency (both dimensionless understood). The imaginary part of these quantities, $\alpha_{i}$ and $\omega_{i}$, denote the spatial and temporal growth rates, respectively. If $\omega_{i}>0$ the interface is considered temporally unstable and developing (initially) sinusoidal waves propagate with the phase velocity $v_{p} = \omega_{r}/\alpha_{r}$.

We solve the underlying generalized complex eigenvalue problem numerically by employing a standard Chebyshev collocation method~\cite{Boomkamp_1997} (full description given in Appendix~\S\ref{app:numerical_method}), thereby adjusting the number of collocation points until convergence is reached.

\subsection{Nonlinear direct numerical simulation}
\label{sec:DNS}
To capture and analyse the development of the flow beyond the linear regime, we use the two-phase flow solver presented in Reference~\cite{ONaraigh_2014} for direct numerical simulation of the full Navier-Stokes equations. This in-house solver is level set method based and uses the continuum surface force (CSF) formulation to model surface tension effects~\cite{Brackbill_1992,Sussman_1998}. In level set formalism, the governing equations read as
\begin{subequations}
	\begin{equation}
		\begin{multlined}[b]
			\rho\left(\frac{\partial\bm{u}}{\partial t} + \bm{u\cdot\nabla u}\right) = \\
			-\bm{\nabla} p
			+\frac{1}{Re_{p}}\bm{\nabla\cdot}\left[\mu\left(\bm{\nabla u} + \bm{\nabla u}^{\mathrm{T}}\right)\right]
			+\rho\frac{1}{Fr^{2}}\bm{\hat{e}}_{x}
			+\frac{1}{We}\delta_{\epsilon}\left(\phi\right)\bm{\hat{n}}\kappa,
		\end{multlined}
		\label{eq:full_NS}
	\end{equation}
	\begin{equation}
		\bm{\nabla\cdot u} = 0,
	\label{eq:continuity}
	\end{equation}
	\begin{equation}
		\frac{\partial \phi}{\partial t} + \bm{u\cdot\nabla}\phi = 0
	\label{eq:advection_levelset}
	\end{equation}
	\begin{equation}
		\bm{\hat{n}} = \frac{\bm{\nabla}\phi}{\left|\bm{\nabla}\phi\right|}, \quad \kappa = -\bm{\nabla\cdot\hat{n}}.
	\label{eq:unit_normal_curvature}
	\end{equation}
	\label{eq:level_set}%
\end{subequations}
Here, $\phi\left(\bm{x},t\right)$ is the level set function indicating the phase in which a point $\bm{x}$ lies (liquid phase for $\phi<0$, gas phase for $\phi>0$). Hence, the zero level set, $\phi\left(\bm{x},t\right) = 0$, represents the interface $\eta\left(x,t\right)$. The level set function also determines the unit vector $\bm{\hat{n}}$ normal to the interface and the interface curvature $\kappa$ in Eq.~\eqref{eq:unit_normal_curvature}. As the level set function differentiates between the two phases, it is also used to identify the respective density and viscosity through the expressions $\rho = H_{\epsilon}\left(\phi\right)+r\left(1-H_{\epsilon}\left(\phi\right)\right)$ and $\mu = H_{\epsilon}\left(\phi\right)+m\left(1-H_{\epsilon}\left(\phi\right)\right)$. The function $H_{\epsilon}\left(\phi\right)$ is a regularised Heaviside function, which is smooth in an interval $\left[-\epsilon,\epsilon\right]$ around the interface with $\epsilon$ set equal to $1.5$ times the grid spacing. This interval also supports the regularised delta function $\delta_{\epsilon}\left(\phi\right) = \mathrm{d}H_{\epsilon}\left(\phi\right)/\mathrm{d}\phi$.

To discretize Eq.~\eqref{eq:level_set}, we use an isotropic MAC grid with a spacing that resolves the undisturbed liquid film with at least 30 grid points. Additionally, all simulations have been checked for convergence. On the implemented grid, vector quantities are defined at cell faces and scalar quantities are defined at the respective cell centres. A third-order Adams-Bashforth scheme is used to treat the convective derivative, while the momentum fluxes are treated in a flux-conservative fashion employing a (semi-implicit) combination of Crank-Nicholson and third-order Adams-Bashforth method~\cite{boyd_2001}. Pressure and the associated incompressibility of the flow are treated using standard projection method~\cite{Chorin_1968}. We use a combination of Jacobi's method and successive over-relaxation on a red-black scheme to evaluate the predictor and corrector step. Moreover, a third-order (fifth-order accurate) WENO scheme~\cite{jiang_1996} is used to advect the level set function $\phi$, which is subsequently reinitialised applying a Hamilton-Jacobi equation and the algorithm formulated in Reference~\cite{Russo_2000}.

At the domain boundaries we apply standard no-slip and no-penetration conditions at the confining walls, $z = 0$ and $z = H$ (note that the coordinate system underlying the Eqs.~\eqref{eq:level_set} is shifted by $-\delta_{L}$ compared to the one shown in Fig.~\ref{fig:schematic_counter_laminar_laminar}), as well as periodicity in $x$-direction. The pressure is decomposed as $p = \tilde{p} + \left(\mathrm{d}P/\mathrm{d}L\right)x$, where $\tilde{p}$ satisfies the periodic boundary conditions in $x$-direction and $\mathrm{d}P/\mathrm{d}L$ is a positive, constant, dimensionless pressure gradient. Solving the standard force balance in both phases (Eq.~\eqref{app:eq:liquid_NS} and~\eqref{app:eq:gas_NS}) numerically gives the initial velocity field (base flow).

The solver is implemented in Fortran 90 using MPI to decompose the computational domain (Fig.~\ref{fig:schematic_counter_laminar_laminar}) in $x$-direction. This parallelization scheme makes efficient use of the architecture of the UK's ``Advanced Research Computing High End Resource'' (ARCHER, \url{http://www.archer.ac.uk}) on which we run our high resolution simulations.

\section{Temporal stability analysis}
\label{sec:temporal_stability}
In this section we restrict ourselves to the study of the temporal evolution of an (initially) infinitesimally small perturbation of the liquid interface. This temporal framework provides deep insight into the onset of interfacial waves and, thus, the complex dynamics of vertical films sheared by counter-current gas flows. It further enables us to map flow regimes typical for such systems.

We analyse the temporal stability for two distinct cases:
\begin{itemize}
\item High density contrast: we assume a liquid density of $\rho_{L}$ = \SI{1000}{\kilogram\per\metre\cubed}, corresponding to a gas-liquid flow typified by an air and water combination.
We demonstrate below that the large density contrast leads to a complicated characterization of the instability, consisting of competing and coalescing linearly unstable modes.
\item Low density contrast: we assume a liquid density of $\rho_{L}$ = \SI{10}{\kilogram\per\metre\cubed}. The low-density-contrast scenario is popular in the
simulation literature~\cite{Scardovelli_1999,Boeck_2007,Fuster_2013} for a number of reasons. In the present context, it corresponds to a system without mode competition. This provides a ``clean'' database of linear stability results, which can be used as an unambiguous benchmark for direct numerical simulations. 
\end{itemize}
For both considered cases, we further assume a liquid dynamic viscosity of $\mu_{L}$ = \SI{500e-6}{\pascal\second} and a surface tension of $\gamma$ = \SI{1e-3}{\newton\per\metre}, while the dynamic viscosity and density on the gas side are $\mu_{G}$ = \SI{10e-6}{\pascal\second} and $\rho_{G}$ = \SI{1}{\kilogram\per\metre\cubed}, respectively. The flow is confined by a channel of the height $H$ = \SI{0.01}{\metre}. These values result in density ratios of $r = 1000$ and $10$, a viscosity ratio of $m = 50$ as well as a gravity Reynolds number of $Re_{g} = 313$ and leaves the relative film thickness $\delta_{L}$ and the Froude number $Fr$, which can be related to liquid and gas flow rates, respectively, as the remaining parameters to determine the two-phase flow.

Throughout the linear stability analysis, i.e. both density ratio cases, we consider $\delta_{L}\in\left[0.02,0.14\right]$, whereas the Froude number is varied within the interval $Fr\in\left[1.05,13\right]$ for the case of high density contrast. This corresponds to an absolute film thickness $d_{L}$ ranging from \SIrange{0.2e-3}{1.4e-3}{\metre} and an applied pressure drop $\mathrm{d}p/\mathrm{d}x$ in the range of \SIrange{10.8}{1657.9}{\pascal\per\metre}. Within this parameter space, we determine the temporal dispersion relation $\omega^{temp} = \omega\left(\alpha_{r},\alpha_{i} = 0\right)$ numerically on a grid with $\Delta\delta_{L} = 0.005$ and $\Delta Fr = 0.25$. The associated eigenvalue problem, Eq.~\eqref{app:eq:eigenvalue_matrix}, is solved for $\alpha_r \in \left[0.05,\alpha_{c}\right]$ with $\Delta\alpha_{r} = 0.05$, where $\alpha_{c}$ is the cut-off wavenumber beyond which $\omega_{i}^{temp} < 0$. In the low-density-contrast case, we apply the interval $Fr\in\left[1.05,1.55\right]$, corresponding to $\mathrm{d}p/\mathrm{d}x\in\left[10.8,23.6\right]\si{\pascal\per\metre}$, with $\Delta Fr = 0.025$ and determine the dispersion relation for $\alpha_r \in \left[0.05,15\right]$ with $\Delta\alpha_{r} = 0.01$. For each parameter set $\left(\delta_{L},Fr\right)$ we further identify the pair $\left(\alpha_{m}^{temp},\omega_{i}^{temp}\right)$ that maximizes $\omega_{i}$ as the linearly most unstable mode.

\begin{figure}[htbp]%
	\centering
		\includegraphics[width=\textwidth]{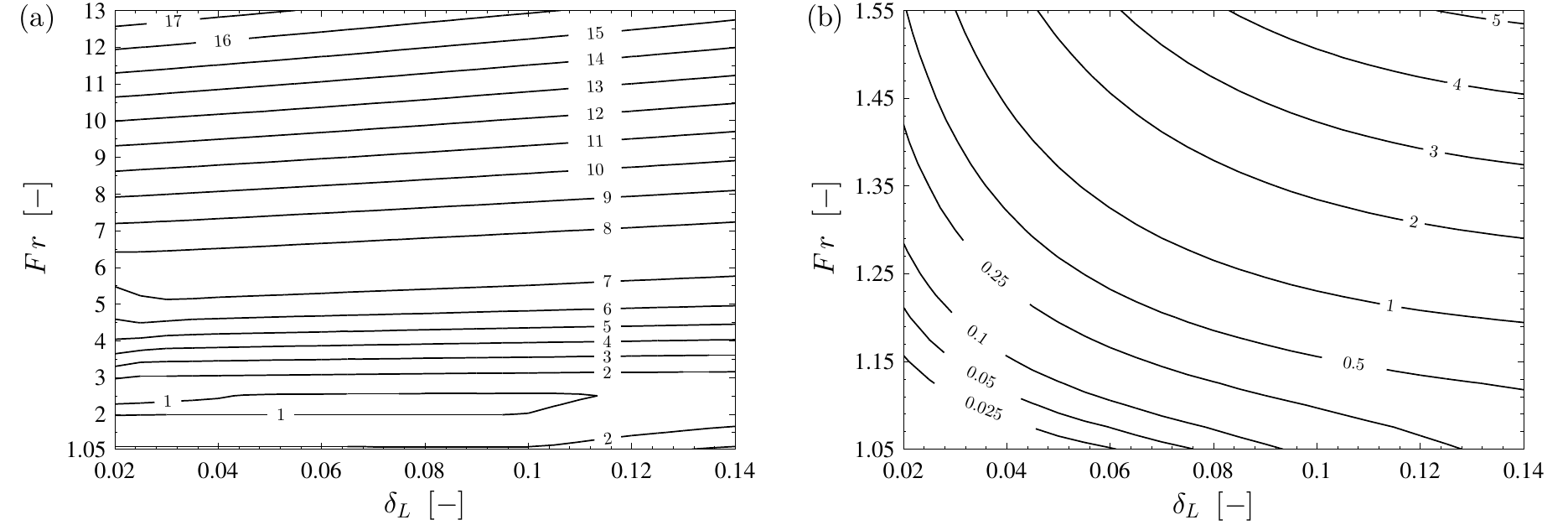}
	\caption{Temporal growth rate $\omega_{i}^{temp}$ of the linearly most unstable mode in the parameter spaces of the two investigated cases. (a) $r=1000$; (b) $r=10$.}
	\label{fig:comparison_temporal_linear_growth_r1000_r10}
\end{figure}

\begin{figure}[htbp]%
	\centering
		\includegraphics[width=\textwidth]{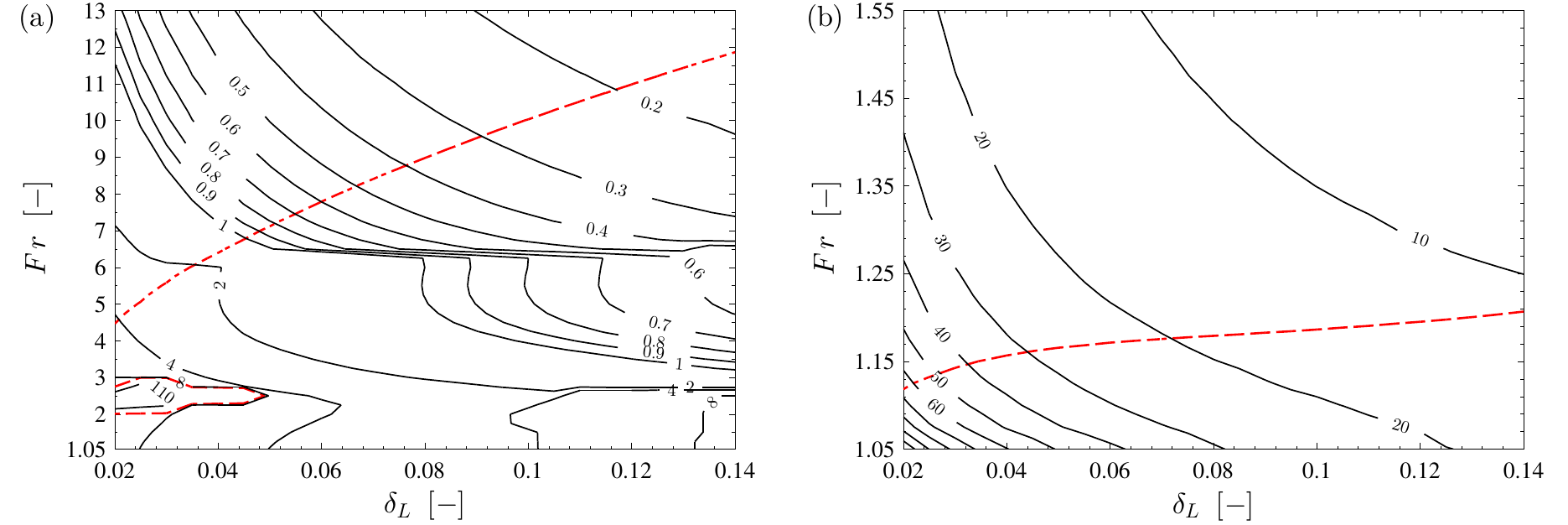}
	\caption{Wavelength $\lambda_{m}$ of the linearly most unstable mode scaled by the corresponding relative film thickness $\delta_{L}$ in the parameter spaces of the two investigated cases. (a) $r=1000$; (b) $r=10$. The dashed line indicates zero phase velocity (onset of flooding) for reference.}
	\label{fig:comparison_wavelength_max_scaled_r1000_r10}
\end{figure}

For both density ratio cases, our analysis shows (Fig.~\ref{fig:comparison_temporal_linear_growth_r1000_r10}) that the temporal growth rate of that mode always attains positive values, which means the interface is inherently unstable within the investigated parameter spaces.
However, the trends for the growth rate of the most unstable mode are quite different in the two cases. Apart from a general increase of $\omega_{i}^{temp}$ with increasing Froude number in the case with high density ratio, a valley stretching out parallel to the horizontal axis at a Froude number of around 2.25 becomes evident in the $\delta_{L}$-$Fr$-plane (Fig.~\ref{fig:comparison_temporal_linear_growth_r1000_r10}a). Moreover, a plateau-like region of nearly equal temporal growth rates appears at about $Fr = 6.00$ spanning all considered film thicknesses. Further insight into the nature of the system can be drawn from the length of the most unstable wave, $\lambda_{m} = 2\pi/\alpha_{m}^{temp}$, and its extent relative to the respective film thickness, $\lambda_{m}/\delta_{L}$. Figure~\ref{fig:comparison_wavelength_max_scaled_r1000_r10}a shows the change of the latter quantity in the parameter space for density ratio $r = 1000$, revealing the complex interfacial dynamics of the flow.
Although the interface is susceptible to a wide range of linearly most unstable waves, short-wave instability is predominant in this case, which is a consequence of the low surface tension considered herein~\cite{Yiantsios_1988}.

\begin{figure}[htbp]%
	\centering
		\includegraphics[width=\textwidth]{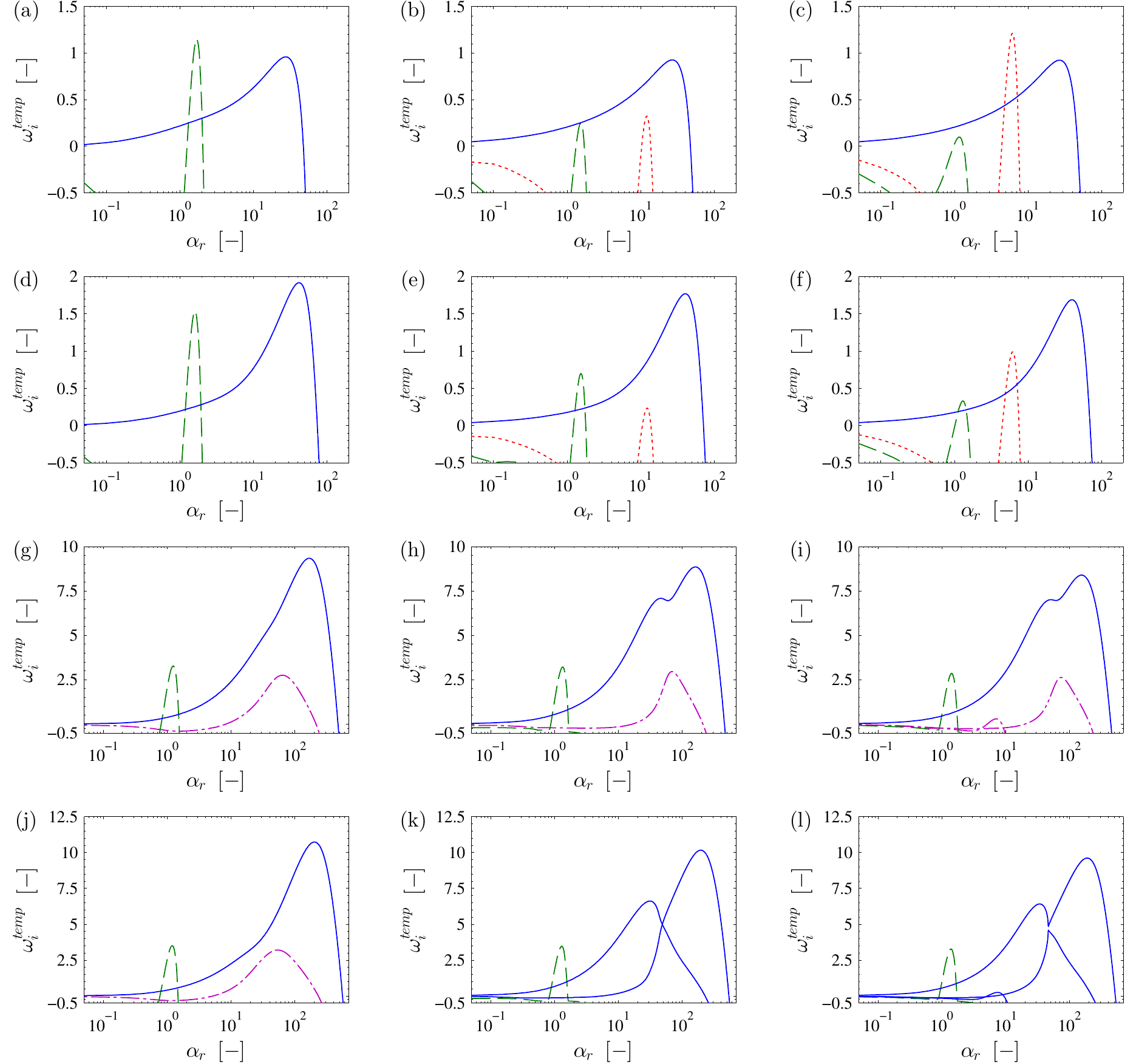}
	\caption{Selected dispersion relations for the case $r = 1000$. Solid lines: interfacial mode; dashed lines: shear mode in gas layer; dotted lines: shear mode in liquid layer; dot-dashed line: internal mode. Top to bottom: $Fr\in\left[2.5,3.0,7.5,8.5\right]$; left to right: $\delta_{L}\in\left[0.03,0.08,0.13\right]$.}
	\label{fig:dispersion_relations_r1000}
\end{figure}

The rich dynamics observed for the high-density-ratio case can be explained by looking at the dispersion relation of representative points in the parameter space (Fig.~\ref{fig:dispersion_relations_r1000}).
It becomes apparent that multiple unstable modes are active, which are coalescing and competing with each other. It is this mode interaction that causes the complex structures presented above. In contrast, the case with low density ratio shows mostly one unstable mode throughout the considered parameter space. Even though further unstable modes are observed, these modes are very weak and appear only occasionally. Hence, for this case the system behaviour is less intricate (Fig.~\ref{fig:comparison_temporal_linear_growth_r1000_r10}b and~\ref{fig:comparison_wavelength_max_scaled_r1000_r10}b).

In order to identify the character of these unstable modes and hence the driving force of the instability in these scenarios, we characterize the energy transfer from the base flow by decomposing the disturbance kinetic energy into production and dissipation terms. However, as both presented parameter spaces are quite extensive, we restrict ourselves to the detailed analysis of the representative scenarios listed in Table~\ref{tab:temporal_scenarios}.
\begin{table}[htbp]
	\begin{ruledtabular}
		\begin{tabular}{ccccccccccc}
			r	& Scenario & $\delta_{L}$	& $Fr$	& $Re_{p}$	& $We$	& $\alpha_{m}^{temp}$	& $v_{p, OS}$	& $\omega^{temp}_{i, OS}$	& $\omega^{temp}_{i, DNS}$\\
			\hline
			\multirow{3}{*}{1000}	& T1C	& 0.13	& 3.00	& 939		& \hphantom{5}8.829	& \hphantom{1}40.05	& 16.47												& 1.6853	& -\\			
																									& T2C	& 0.08	& 7.50	& 2349	& 55.181												& 161.90												& \hphantom{1}0.82	& 8.8447	& -\\
																									& T3C	& 0.08	& 3.00	& 939		& \hphantom{5}8.829	& \hphantom{1}40.85	& \hphantom{1}5.97	& 1.7662	& 1.7783\\
			\hline
			\multirow{4}{*}{10}			& dT4C	& \multirow{4}{*}{0.08}	& 1.157	& 362	& 1.313	& 3.99	& \hphantom{-}0.05	& 0.3662	& 0.3668\\
																									& dT5L		&																						& 1.179	& 369	& 1.363	& 4.29	& \hphantom{-}0.00	& 0.4669	& 0.4634\\
																									& dT6F		&																						& 1.201	& 376	& 1.415	& 4.59	& -0.05											& 0.5829	& 0.5790\\
																									& dT7F		&																						& 1.319	& 413	& 1.706	& 6.19	& -0.30											& 1.4347	& 1.3806\\
		\end{tabular}
	\end{ruledtabular}
	\caption{Conditions studied in detail using linear theory and nonlinear direct numerical simulations (case $r = 10$ only). The leading letter of the scenario name designates temporal stability analysis (prefix ``d'' for decreased density ratio), followed by a running number for this type of analysis; the trailing letter corresponds to the flow regime (\textbf{C}ounter-current, \textbf{L}oading, \textbf{F}looding; introduced in \S~\ref{sec:flow_regimes}) in which the pair $\left(\delta_L, Fr\right)$ is situated.}
	\label{tab:temporal_scenarios}
\end{table}
The mode characteristics for all other parameter sets are deduced by inspection.

\subsection{Energy budget}
\label{sec:energy_budget}
Following the approach in Reference~\cite{Boomkamp_1996}, the rate of change of kinetic energy in both phases $KIN_{L,G}$ can be decomposed into
\begin{equation}
	KIN_{L} + KIN_{G} = DISS_{L} + DISS_{G} + REY_{L} + REY_{G} + NOR + TAN,
\label{eq:energy_budget}
\end{equation}
where all terms are spatially averaged contributions of the respective fluid to the overall energy budget of the disturbance. The terms $DISS_{L,G}$ represent energy losses due to viscous dissipation, whereas $REY_{L,G}$ denotes wave-induced Reynolds stresses transferring energy between the base flow and the perturbation in the bulk of the two phases. Lastly, $NOR$ and $TAN$ describe the work done by normal and tangential stresses at the interface. The energy budgets for the investigated scenarios are given in Table~\ref{tab:energy_budget}, where the individual terms have been scaled by the total rate of change of kinetic energy ($KIN_{L} + KIN_{G} = 1$).
\begin{table}[htbp]
	\begin{ruledtabular}
		\begin{tabular}{ccccccccccc}
			r	& Scenario	& $\alpha$	& $KIN_{L}$	& $KIN_{G}$	& $DISS_{L}$	& $DISS_{G}$	& $REY_{L}$	& $REY_{G}$	& $NOR$	& $TAN$\\
			\hline
			\multirow{4}{*}{1000}	& T1C		& \hphantom{4}1.30		& 0.02	& 0.98	& -0.01	& \hphantom{3}-1.39	&-0.04
 & \hphantom{-}2.42	& \hphantom{-}0.00	& \hphantom{3}0.02\\
																									& T1C		& \hphantom{4}6.20		& 1.00	& 0.00	& -0.33	& \hphantom{3}-0.01	& \hphantom{-}1.31	& \hphantom{-}0.02	& \hphantom{-}0.00	& \hphantom{3}0.01\\
																									& T1C		& 40.05													& 0.51	& 0.49	& -0.46	& -31.37												& -0.05
 & -4.82											& -0.30											& 37.98\\
																									& T2C		& 69.50													& 0.76	& 0.24	& -0.31	& -18.65												& \hphantom{-}1.06	& -3.67											& -0.02											& 22.57\\
			\hline
			\multirow{4}{*}{10}			& dT4C	& 3.99														& 0.06	& 0.94	& -1.09	& -13.93												& \hphantom{-}0.00	& -1.77											& -0.46											& 18.25\\
																									& dT5L		& 4.29														& 0.09	& 0.91	& -1.21	& -12.24												& \hphantom{-}0.00	& -1.19											& -0.44											& 16.08\\
																									& dT6F		& 4.59														& 0.12	& 0.88	& -1.36	& -10.67												& \hphantom{-}0.00	& -0.72											& -0.42											& 14.17\\
																									& dT7F		& 6.19														& 0.32	& 0.68	& -1.48	& \hphantom{1}-5.44	& \hphantom{-}0.00	& \hphantom{-}0.44	& -0.32											& \hphantom{1}7.79\\
		\end{tabular}
	\end{ruledtabular}
	\caption{Energy budgets of the active modes for the scenarios listed in Table~\ref{tab:temporal_scenarios}.}
	\label{tab:energy_budget}
\end{table}

Decomposition of the energy budget in scenario T1C (Fig.~\ref{fig:dispersion_relations_r1000}f) uncovers three unstable (active) modes: one interfacial mode as well as one shear mode in each fluid layer. The first mode $\left(\alpha = 40.05\right)$ originates from the viscosity and density contrast of the two fluids, which give rise to velocity and stress disturbances tangential to the liquid interface driving the instability (``viscosity-gravity-induced instability''). Across the entire parameter space of the high-density-ratio case, this interfacial mode is active (Fig.~\ref{fig:dispersion_relations_r1000}) and, generally, its maximum growth rate $\omega_{i}^{temp}$ as well as the associated wavenumber $\alpha_{m}^{temp}$ increase for increasing Froude number yet slightly decrease with increasing relative film thickness, see Fig.~\ref{fig:comparison_temporal_linear_growth_r1000_r10}a and~\ref{fig:comparison_wavelength_max_scaled_r1000_r10}a.

The shear mode, or Tollmien-Schlichting mode, in the gas layer is also active for all values of $\delta_{L}$ above a minimum Froude number ranging from \numrange{1.46}{2.34} for increasing film thickness. Above this threshold, the mode becomes stronger with increasing gas flow, though thicker films exhibit slightly lower growth rates. The opposite trend is observed for the wavenumber at maximum growth rate, which shifts towards lower values with increasing $Fr$ and generally higher values for thicker films. Contrasting the behaviour of the shear mode in the gas layer, the liquid layer shear mode is only active for thick enough films $\left(\delta_{L} > 0.067\right)$. While this threshold value increases with increasing Froude number, the maximum growth rate of this mode drops, thereby rendering it less unstable. Further complexity is added to the system dynamics by the fact that the energy contribution to this particular mode due to tangential stresses at the interface becomes larger for increasing $Fr$ and reaches significant values $\left(TAN \approx 1\right)$. Eventually, $TAN$ becomes the dominant term but with $REY_{L}$ still being of importance to overcome the restoring effects in both phase. This combination of energy sources driving the instability is also known as internal mode.

Apart from this change in mode identity, there is a further internal mode emerging at Froude numbers ranging from \numrange{5.21}{5.55} for increasing $\delta_{L}$, e.g. scenario T2C or Fig.~\ref{fig:dispersion_relations_r1000}g. As this mode grows stronger with higher gas flow rates, the wavenumber $\alpha_{m}^{temp}$ at its maximum growth rate shift towards lower values. Furthermore, the mode starts to coalesce with the dominant interfacial mode, which leads to the formation of a second ``hump'' in that dispersion relation (Fig.~\ref{fig:dispersion_relations_r1000}h). For even higher $Fr$, the coalescing internal mode undergoes a change in identity itself and forms a second active interfacial mode (Fig.~\ref{fig:dispersion_relations_r1000}k and~\ref{fig:dispersion_relations_r1000}l). This ``splitting'' of the dispersion relation induces a region of rapidly changing wavelength of the linearly most unstable mode at intermediate Froude numbers (Fig.~\ref{fig:comparison_wavelength_max_scaled_r1000_r10}a). Although mode coalescence occurs throughout the entire range of investigated film thicknesses, splitting of the dispersion relation is not observed for low values of $\delta_{L}$ (Fig.~\ref{fig:dispersion_relations_r1000}j). It is further worth mentioning that additional modes can become unstable, which are, however, mostly temporarily active and thus of minor significance.

The multitude of active and coexisting modes leads not only to mode coalescence but also to a certain amount of mode competition. In the high-density-contrast case this mode competition mainly occurs at low Froude numbers, where the growth rate associated with the interfacial mode is still comparatively low. The shear modes at low and high values of $\delta_{L}$, on the other hand, exhibit stronger growth and therefore constitute the dominant mode (Fig.~\ref{fig:dispersion_relations_r1000}a and~\ref{fig:dispersion_relations_r1000}c). As $Fr$ increases, the interfacial mode picks up strength and supersedes the shear modes as the dominating mode (Fig.~\ref{fig:dispersion_relations_r1000}d and~\ref{fig:dispersion_relations_r1000}f), which results in a jump of the wavenumber $\alpha_{m}^{temp}$ associated with the maximum growth rate. This jump explains the regions of relatively long waves at low and high values of $\delta_{L}$ in Fig.~\ref{fig:comparison_wavelength_max_scaled_r1000_r10}a.

As mentioned above, the low density-ratio-case exhibits mainly one unstable mode, which is due to the density and viscosity contrast of the two fluids (Table~\ref{tab:energy_budget}). Although both differences account for energy transferred towards the disturbed flow, the contribution related to the viscosity contrast dominates. Hence, this mechanism is, in general, consistent with the so-called Yih mode~\cite{Yih_1967}. It is further apparent that the relative fraction of kinetic energy associated with the liquid phase increases with increasing Froude number. This rise, together with an enhanced energy dissipation, can be linked to more agitation in the liquid film as we will show in section~\ref{sec:nonlinear_wave_dynamics}.
The amount of energy dissipated in the gas phase, on the other hand, seems to drop, which is counter-intuitive for an increased $Fr$. Yet, dissipation in the gas does increase in absolute terms but at a slower rate as the total kinetic energy. That, in turn, leads to the seemingly decreasing rate of dissipation in the gas phase. The same effect can be seen for $NOR$ and $TAN$.

Another result worth noting is the change in sign of $REY_{G}$ as $Fr$ increases, turning wave-induced Reynolds stresses from an additional dissipative energy ``sink'' to an energy ``source''. This, in turn, can be explained by an unstable Tollmien-Schlichting mode appearing in the gas stream (scenario dT7F) that delivers energy to the interfacial instability, thereby suggesting a transition to turbulence in the bulk of the gas phase. Therefore, this particular scenario can conceptually not be regarded as a ``pure'' Yih-type instability any more but tends towards an internal mode. This positive contribution of wave-induced Reynolds stresses to the instability occurs throughout the entire parameter space but is shifted to lower Froude numbers for thicker liquid films. However, it has to be emphasized that the tangential stresses doing work at the interface are the dominant driving force of the instability in all presented scenarios for the low-density-ratio case (Table~\ref{tab:energy_budget}). Across the entire parameter space the maximum growth rate $\omega_{i}^{temp}$ of this interfacial mode increases with increasing $\delta_{L}$ and $Fr$ (Fig.~\ref{fig:comparison_temporal_linear_growth_r1000_r10}b), whereas the associated wavenumber $\alpha_{m}^{temp}$ increases with increasing Froude number but decreases for thicker films, which in agreement with~\cite{Dietze_2013}. Overall, the system is predominantly receptive to long-wave instability under low-density-ratio conditions (Fig.~\ref{fig:comparison_wavelength_max_scaled_r1000_r10}b).

\subsection{Flow regimes}
\label{sec:flow_regimes}
\begin{figure}[htbp]%
	\centering
		\includegraphics[width=\textwidth]{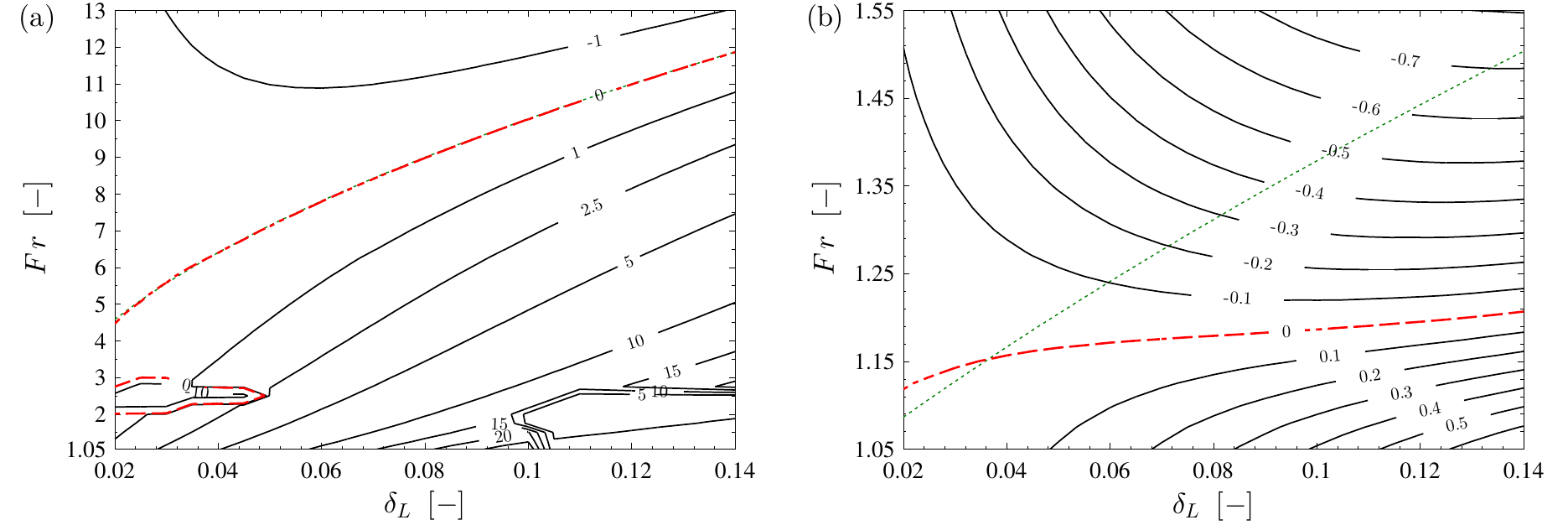}
	\caption{Phase velocity $v_{p}$ of the linearly most unstable mode in the parameter spaces of the two investigated cases. (a) $r=1000$; (b) $r=10$. Zero phase velocity (dashed line) marks the onset of flooding. For comparison, the curve of zero interfacial velocity in the undisturbed base flow (dotted line).}
	\label{fig:comparison_c_p_loading_curve_r1000_r10}
\end{figure}

Fig.~\ref{fig:comparison_c_p_loading_curve_r1000_r10} shows the phase velocity $v_{p}$ of the fastest growing wave developing on the interface for both density ratio cases.
It becomes apparent that the parameter space is divided into two regimes: one in which developing waves exhibit a positive phase velocity and another in which the phase velocity is negative. With respect to the chosen coordinate system (Fig.~\ref{fig:schematic_counter_laminar_laminar}), these regions correspond to waves propagating downwards and upwards, respectively. The vanishing $v_{p}$ at the demarcation (dashed lines in Fig.~\ref{fig:comparison_c_p_loading_curve_r1000_r10}) between these regimes relates therefore to a standing wave. By crossing this line with increasing $Fr$, liquid adjacent to the interface begins to move upwards with the developing wave. This (partial) upward flow of the liquid phase is considered as \emph{flooding}~\cite{Bankoff_1986}. The curve $v_{p}\left(\delta_{L},Fr\right) = 0$, which is referred to as \emph{loading curve} in chemical engineering, indicates the operational limit of vertical counter-current gas-liquid flows. Thus, the region below the loading curve constitutes the counter-current regime.

The flow map for the case with high density ratio (Fig.~\ref{fig:comparison_c_p_loading_curve_r1000_r10}a) reflects the complex interfacial dynamics described above. In regions with mode competition and changing dominant mode, the phase velocity changes at a great rate due to the jump of the wavenumber $\alpha_{m}^{temp}$ associated with the linearly most unstable mode. For low Froude numbers, this leads to a region of flooding amidst the counter-current regime at the thin film end of the parameter space, while a region of comparatively slow waves occurs at the high film thickness end. In contrast, the phase velocity changes smoothly in the low-density-ratio case due to the absence of mode competition (Fig.~\ref{fig:comparison_c_p_loading_curve_r1000_r10}b).

To illustrate the effect of the linearly most unstable waves on the onset of flooding, Fig.~\ref{fig:comparison_c_p_loading_curve_r1000_r10} also shows the curve of zero interfacial velocity (dotted line) for the undisturbed base flow of the respective case, which can be seen as the analogue to the actual loading curve. In the high-density-ratio case the two curves (with exception of the ``island'' region) almost coincide, i.e. linear waves have no significant influence on the loading curve (Fig.~\ref{fig:comparison_c_p_loading_curve_r1000_r10}a). This can be explained by looking at the expression for the phase velocity, which can be decomposed (trivially) as $v_{p} = u_{0,int} + u_1(\alpha)$, where $u_{0,int}$ denotes the interfacial velocity of the undisturbed base flow. Yet, a good approximation for $u_1(\alpha)$ can be found in the dispersion relation for free-surface capillary waves~\cite{ONaraigh_2011,Miles_1957}, such that $u_1(\alpha)\approx \omega_{cap}/\alpha$, where $\omega_{cap}^{2} = \gamma\alpha^{3}/\left(\rho_{L}+\rho_{G}\right)$ (dimensional understood). Even in the absence of this particular approximation, it can be argued that $u_1(\alpha)$ should be symmetric in the densities (at least for the present vertical configuration), and should be a monotonically decreasing function of the densities, as generally the inertia tends to affect the phase speed in a reciprocal manner, i.e. the larger the inertia the smaller the phase speed. Thus, it can be seen that that $v_{p}\rightarrow u_{0,int}$ for high density ratio. This also implies that interfacial waves do alter the flooding curve in the case with low density ratio. The counter-current regime is significantly reduced by the instability for large parts of the parameter space (Fig.~\ref{fig:comparison_c_p_loading_curve_r1000_r10}b). However, below a threshold of $\delta_{L} = 0.036$ the onset of flooding is pushed towards higher Froude numbers.

\section{Absolute and convective instability of the linear dynamics}
\label{sec:A/C_instability}
Complementing the temporal analysis of the interfacial instability, we study the response of the system to a pulse on the liquid interface. In particular, we are interested whether this initially localized disturbance spreads both up- and downstream and eventually contaminates the entire flow, rendering the system \emph{absolutely unstable}. In case the disturbance propagates away from its source, the flow is considered \emph{convectively unstable}.

To determine the nature of the instability in this context, the complex dispersion relation $D\left(\alpha,\omega\right) = 0$, which is obtained by solving the eigenvalue problem of Eq.~\eqref{app:eq:eigenvalue_operator} numerically for a range of complex wavenumbers $\alpha = \alpha_{r} + \mathrm{i}\alpha_{i}$, has to be evaluated against conditions essential for absolute instability as outlined in Reference~\cite{Huerre_1990}: (i) a positive imaginary part of the angular frequency $\omega_{i,0} \coloneqq \omega_{i}\left(\alpha_{S}\right) > 0$ at a saddle point $\alpha_{S}$ in the complex $\alpha$-plane, where $\alpha_{S}$ solves $\mathrm{d}\omega/\mathrm{d}\alpha = 0$, forms the necessary condition; (ii) to satisfy the sufficient condition, spatial branches $\alpha^{\pm}\left(\omega\right)$ that originate from opposite halves of the $\alpha$-plane have to coalesce at the saddle point $\alpha_{S}$, forming a genuine pinch point (this coalescence corresponds to the formation of a branch point/cusp at $\omega_{i,0}$ in the complex $\omega$-plane~\cite{Kupfer_1987}). Meeting both conditions will result in growth of the disturbance at its source with the \emph{absolute growth rate} $\omega_{i,0}$.

\begin{figure}[htbp]
	\centering
		\includegraphics[scale=1.00]{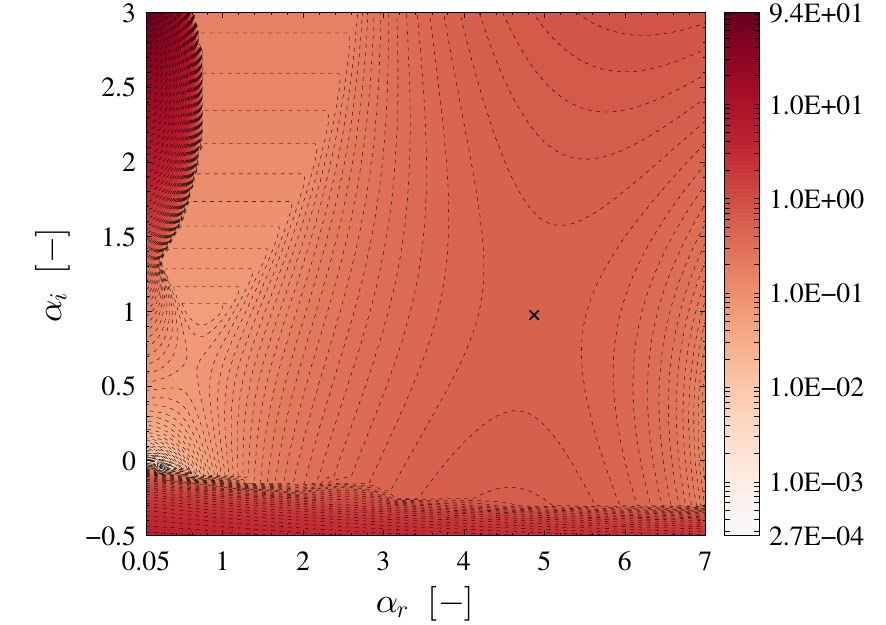}
	\caption{Global topography of $\omega_{i}$ for scenario ST4 ($r = 10$, $\delta_{L} = 0.08$, $Fr = 1.201$, $Re_{p} = 376$, $We = 1.415$) as obtained from spatio-temporal Orr-Sommerfeld analysis. The cross marks the pinching saddle point.}
	\label{fig:contour_omega_i_spatio_temporal}
\end{figure}

Although the described procedure seems straightforward, inspection of the results of a spatio-temporal Orr-Sommerfeld (ST-OS) analysis requires great care in order to avoid misinterpretation due to the complexity of the dispersion relation that can arise from the multivalued nature of the eigenvalue problem as well as specifics of the investigated problem, such as applied boundary conditions or multiple unstable temporal modes (see Fig.~\ref{fig:contour_omega_i_spatio_temporal}).
Hence, to conclude the character of the instability correctly, it is necessary to examine the global topography of frequency over wavenumber for each relevant set of parameters. As we are interested in identifying convective/absolute instability (C/A) transition, the described method is not practical given the large parameter space considered herein. Instead, we make use of the analytical connection between temporal and spatio-temporal frequency presented in Reference~\cite{ONaraigh_2013}.

The authors derived an expression for $\omega_{i}\left(\alpha_{r},\alpha_{i}\right)$ based on analytical continuation of purely temporal stability properties, specified on the real axis, into the complex $\alpha$-plane. Accounting for up to second-order terms, the quadratic approximation (QA) of the imaginary part of the complex frequency reads
\begin{equation}
	\omega_{i}\left(\alpha_{r},\alpha_{i}\right) =
			\omega_{i}^{temp}\left(\alpha_{r}\right) +
			\frac{\mathrm{d}\omega_{r}^{temp}\left(\alpha_{r}\right)}{\mathrm{d}\alpha_{r}}\alpha_{i} -
			\frac{1}{2}\frac{\mathrm{d}^{2}\omega_{i}^{temp}\left(\alpha_{r}\right)}{\mathrm{d}\alpha_{r}^{2}}\alpha_{i}^{2}.
\label{eq:quad_approx}
\end{equation}
Similarly, we approximate $\omega^{temp}\left(\alpha_{r}\right)$ in the vicinity of the temporally most unstable mode by second-order polynomials. With the conditions $\partial\omega_{i}/\partial\alpha_{r} = \partial\omega_{i}/\partial\alpha_{i} = 0$ applied to Eq.~\eqref{eq:quad_approx}, we can readily estimate the position of the saddle point $\alpha_{S}$ and the absolute growth rate $\omega_{i,0}$ for a given set of flow parameters, requiring results from temporal linear stability analysis only. Even though this approach seems to circumvent the pitfalls associated with the full spatio-temporal linear stability analysis outlined above, one has to keep in mind that the quadratic approximation is, strictly speaking, only valid inside a disc of convergence with radius $R$, where $R$ is the minimum distance from the centre of the complex Taylor series $\left(\alpha_{r} = \alpha_{m}^{temp},0\right)$ to the nearest singularity of $\omega\left(\alpha\right)$.

Like in the purely temporal framework, we study several scenarios in more detail (Table~\ref{tab:spatio_temporal_scenarios}).
\begin{table}[htbp]%
	\begin{ruledtabular}
		\begin{tabular}{ccccccccc}
			Scenario					&
			$\delta_{L}$	&
			$Fr$								&
			$Re_{p}$			&
			$We$																			& Method	& $\alpha_{r,S}$	& $\alpha_{i,S}$			& $\omega_{i,0}$\\
			\hline
			\multirow{2}{*}{ST1}				&
			\multirow{2}{*}{0.08}				&
			\multirow{2}{*}{1.075}			&
			\multirow{2}{*}{337}				&
			\multirow{2}{*}{1.134}			& QA					& 3.34										& -2.24											& -0.0877\\
				&	&	&	&																& DNS				& -													& -															& \hphantom{-}$<$0\\
			\hline
			\multirow{2}{*}{ST2}				&
			\multirow{2}{*}{0.08}				&
			\multirow{2}{*}{1.157}			&
			\multirow{2}{*}{362}				&
			\multirow{2}{*}{1.313}			& QA					& 4.01										& \hphantom{-}0.24	& \hphantom{-}0.3632\\
				&	&	&	&																& ST-OS		& 4.02										& \hphantom{-}0.24	& \hphantom{-}0.3632\\
			\hline
			\multirow{2}{*}{ST3}				&
			\multirow{2}{*}{0.08}				&
			\multirow{2}{*}{1.179}			&
			\multirow{2}{*}{369}				&
			\multirow{2}{*}{1.363}			& QA					& 4.35										& \hphantom{-}0.66	& \hphantom{-}0.4422\\
				&	&	&	&																& ST-OS		& 4.43										& \hphantom{-}0.65	& \hphantom{-}0.4423\\
			\hline
			\multirow{3}{*}{ST4}				&
			\multirow{3}{*}{0.08}				&
			\multirow{3}{*}{1.201}			&
			\multirow{3}{*}{376}				&
			\multirow{3}{*}{1.415}			& QA					& 4.72										& \hphantom{-}1.02	& \hphantom{-}0.5205\\
				&	&	&	&																& ST-OS		& 4.87										& \hphantom{-}0.97	& \hphantom{-}0.5221\\
				&	&	&	&																& DNS				& -													& -															& \hphantom{-}0.5269\\
		\end{tabular}
	\end{ruledtabular}
	\caption{Scenarios with low density ratio ($r = 10)$ studied in the spatio-temporal (ST) framework.}
	\label{tab:spatio_temporal_scenarios}
\end{table}
Using spatio-temporal Orr-Sommerfeld analysis and quadratic approximation as outlined above, we determine the saddle point $\alpha_{S}$ as well as the corresponding absolute growth rate $\omega_{i,0}$ for all listed scenarios and further perform direct numerical simulations for the scenarios ST1 and ST4. In the following, we want to compare and discuss the obtained results for scenario ST4 in more detail.

\begin{figure}[htbp]
	\centering
		\includegraphics[width=\textwidth]{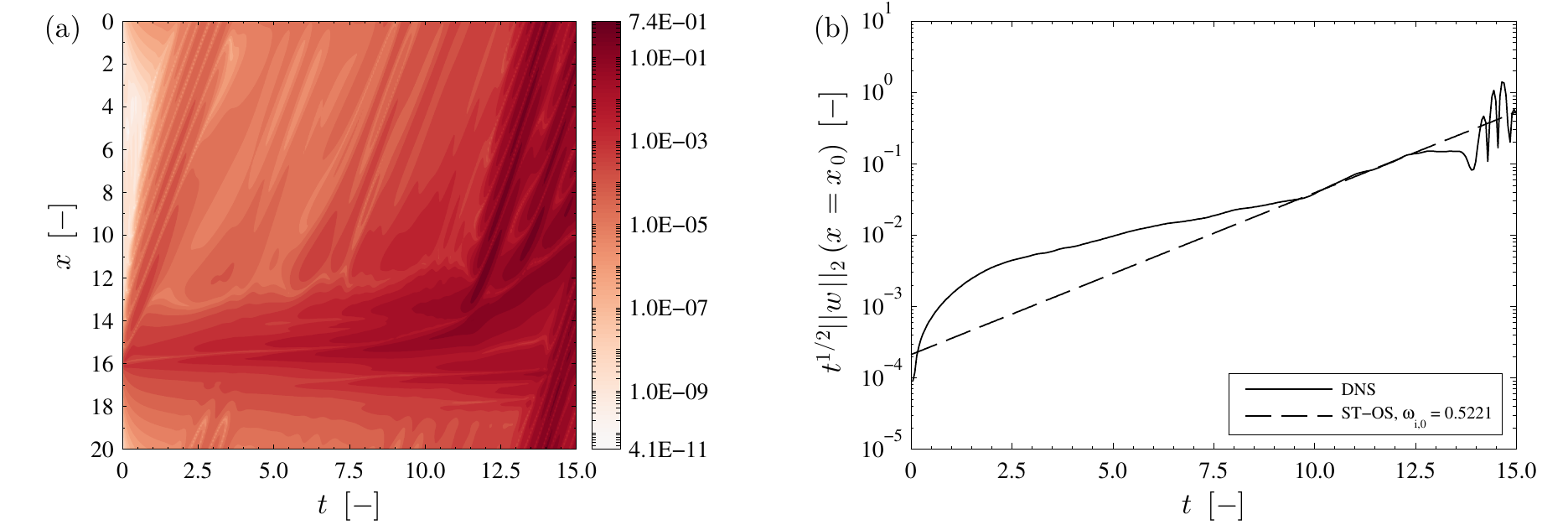}
	\caption{(a) Space-time plot of the norm $n\left(x,t\right)$ for scenario ST4; (b) Comparison between growth rate at the source of the disturbance obtained from DNS and spatio-temporal Orr-Sommerfeld analysis.}
	\label{fig:spatio_temporal_DNS}
\end{figure}

As shown earlier already, the global topography of $\omega_{i}$ in the complex $\alpha$-plane (Fig.~\ref{fig:contour_omega_i_spatio_temporal}) is rather complex. First, the dispersion relation contains two saddle points, of which both may be dynamically relevant. The confinement of the flow by the channel walls has, furthermore, created a discrete pole on the imaginary axis~\cite{ONaraigh_2013}, $\left(\alpha_{r},\alpha_{i}\right) = \left(0,3.34\right)$, which has implications on the character of the saddle point closer to that particular singularity. Lastly, the multivalued nature of the dispersion relation becomes apparent by the branch cut just below the real axis. Although these features make the final characterisation of the instability more difficult, the saddle point at $\left(\alpha_{r},\alpha_{i}\right) = \left(4.87,0.97\right)$ clearly appears as a result of the coalescence of spatial branches emanating from opposite half-planes. It is therefore also pinch point and contributes to spatio-temporal growth at a rate of $\omega_{i,0} = 0.5221$. The positive value of $\alpha_{i}$ indicates that the disturbance travels upwards, which is confirmed by DNS (Fig.~\ref{fig:spatio_temporal_DNS}a). On the other hand, the saddle point at $\left(\alpha_{r},\alpha_{i}\right) = \left(0.62,0.65\right)$ is not a pinch point. In fact, the spatial branch $\alpha^{+}\left(\omega\right)$ (above the saddle point) is a closed curve, whereas the $\alpha^{-}\left(\omega\right)$ branch does not (entirely) originate from the negative half-plane. Hence, this saddle point is dynamically irrelevant regarding absolute instability.

As mentioned before, the singularity closest to the position of the temporally most unstable mode $\left(\alpha_{r} = \alpha_{m}^{temp},0\right)$ determines the disc of convergence in which the quadratic approximation can be applied with confidence. In scenario ST4, the confinement pole at $\left(\alpha_{r},\alpha_{i}\right) = \left(0,3.34\right)$ limits the outermost radius $R$ of this disc to about 5.68. Equation~\eqref{eq:quad_approx} is thus convergent across the relevant section of the complex $\alpha$-plane depicted in Fig.~\ref{fig:contour_omega_i_spatio_temporal}. The approximated position of the pinching saddle point as well as the corresponding growth rate agree well with spatio-temporal OS analysis, see Table~\ref{tab:spatio_temporal_scenarios}.

We further compare the impulse response for the parameters of scenario ST4 captured by DNS against linear theory (ST-OS). In contrast to simulations within the purely temporal framework, a Gaussian pulse with a height of $1\cdot 10^{-3}$ and a standard deviation of $1\cdot 10^{-1}$ is implemented on the interface as an initial condition to study the spatio-temporal nature of the flow. Growth of this perturbation at its source then constitutes absolute instability. However, the developing disturbances will inevitably contaminate the pulse source due to the implemented streamwise periodic boundary conditions. To delay this contamination sufficiently, the channel length is set to $L_{x} = 20$.
Using the pulse norm
\begin{equation*}
	n\left(x,t\right) = \left(\int_{0}^{1}{\left|w\left(x,z,t\right)\right|^{2}\,\mathrm{d}z}\right)^{1/2},
\label{eq:pulse_norm}
\end{equation*}
the space-time plot in Fig.~\ref{fig:spatio_temporal_DNS}a shows the temporal evolution of the perturbations along the streamwise coordinate.
It can be seen that, starting from its initial position at $x = 16$, the pulse is convected upwards, which is in accordance with the results from linear theory mentioned above. This upward motion triggers a pressure disturbance moving ahead of the pulse. However, this ``shock'' decays sufficiently before it re-enters the computational domain, thus avoiding early contamination of the instability source (see left-hand side of Fig.~\ref{fig:spatio_temporal_DNS}a). Further information from this plot is extracted in Fig.~\ref{fig:spatio_temporal_DNS}b, where the growth of the instability at its source is given. After an initial transient period, the growth rate of the instability stabilises and follows the value predicted by ST-OS and QA closely, therefore validating the DNS results. Thereafter, a second pressure shock develops $\left(t\approx11.5\right)$ and travels upwards to eventually contaminate the pulse source at around $t = 14.0$.

The excellent agreement that has been established between linear theory (ST-OS) and quadratic approximation in ST4 can also be observed for the scenarios ST2 and ST3 (Table~\ref{tab:spatio_temporal_scenarios}). For scenario ST1 we were not able to determine the absolute growth rate of the system by means of OS analysis and saddle point method due to the multivalued nature of the associated eigenvalue problem. In fact, the dynamically relevant saddle point appears in the lower half-plane below the branch cut arising near the real axis. To analyse this saddle point a laborious reconstruction of the corresponding part of the Riemann surface from discrete eigenvalues would have to be carried out. However, we avoid this procedure by using the quadratic approximation, which indicates scenario ST1 to be convectively unstable. A further direct numerical simulation confirms this (not shown).

\begin{figure}[htbp]
	\centering
		\includegraphics[scale=1.00]{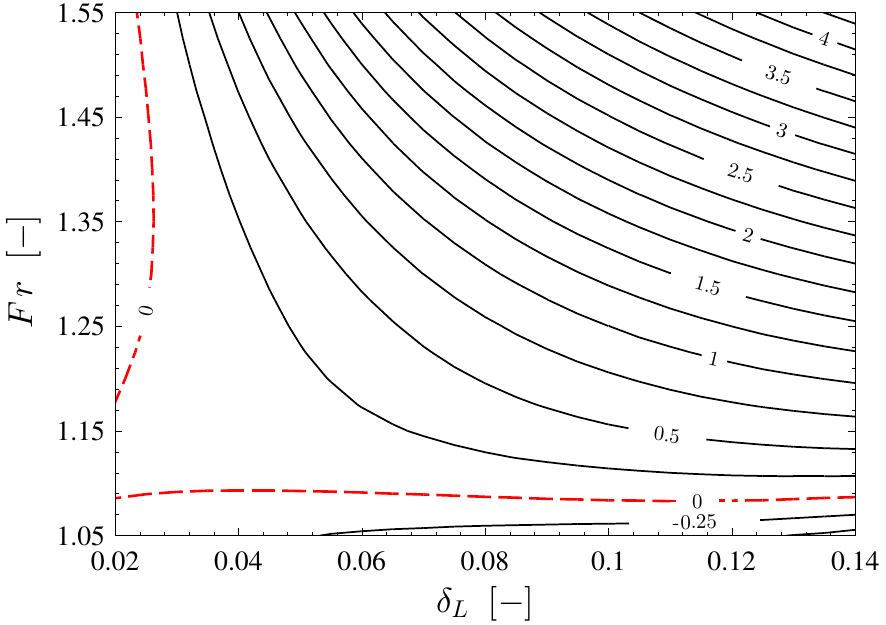}
	\caption{Absolute growth rate $\omega_{i,0}$ at the saddle point (contours) as obtained by analytic continuation (quadratic approximation) from purely temporal quantities. The dashed lines demarcate the transition between convective and absolute instability.}
	\label{fig:contour_omega_i_spatio}
\end{figure}

Given these results, we apply the QA to identify the spatio-temporal nature of the system throughout the wide parameter space considered in the low-density-ratio case. The calculated growth rates $\omega_{i,0}$ are shown in Fig.~\ref{fig:contour_omega_i_spatio}, where the dashed lines demarcate the C/A boundaries.
It becomes apparent that the system is absolutely unstable in almost the entire domain with exception of two narrow bands at the low Froude number and low film thickness end of the parameter range, respectively. In view of these results and those of \S~\ref{sec:flow_regimes}, it appears that C/A transition and the onset of flooding are not closely related for the case of density ratio $r = 10$.

Regarding the accuracy of these results, it has to be mentioned that the relative error $\left(\omega_{\mathrm{QA}}-\omega_{\operatorname{ST-OS}}\right)/\left|\omega_{\operatorname{ST-OS}}\right|$ between QA and linear theory increases with decreasing film thickness and increasing Froude number. At the same time, the quadratic approximation generally underestimates the ``true'' value of $\omega_{i.0}$ as obtained by ST-OS, which leads to a larger regime of absolute instability (towards thin films) than displayed in the above figure.

Finally, the mode coalescence and mode competition encountered in the case with high density ratio (e.g. Figure~\ref{fig:dispersion_relations_r1000}) makes it difficult to identify the nature of the instability in a spatio-temporal framework with both the present methods. To study this manifestation of the instability, alternative techniques, such as linearized DNS~\cite{ONaraigh_2013a} (wherein the linearized equations of motion are solved as a Cauchy problem, but still within the framework the Orr-Sommerfeld linear operators), might be more suitable. A detailed investigation of this particular regime, i.e. high density contrast, will be left to future work. 


\section{Nonlinear wave dynamics}
\label{sec:nonlinear_wave_dynamics}
As temporal linear stability analysis permits the analysis of infinitesimally small interface perturbations only, we carry out direct numerical simulations to study the evolution of these disturbances up to finite amplitudes for both low and high density contrasts. To that end, we use the in-house solver described in \S~\ref{sec:DNS} with streamwise-periodic boundary conditions and initial interface location
\begin{equation}
	\eta\left(x,t=0\right) = \delta_{L} + A_{0}\sum_{n=1}^N\cos\left(\alpha_{n}x+\varphi_{n}\right),
\label{eq:initial_interface}
\end{equation}
where $A_{0}$ is some small initial amplitude (herein $A_{0}$ = \SI{1e-3}), $N$ is the number of linearly unstable modes initialized, $\alpha_{n} = n\left(2\pi/L_{x}\right)n$ is a wavenumber in streamwise direction and $\varphi_{n}$ is a random phase. Even though periodic boundary conditions do not reflect the behaviour of a real system, it is appropriate to consider this setup as it allows for a rigorous comparison of DNS results with linear theory as well as unambiguous results of the Fourier transform taken of the interfacial wave at finite times~\cite{ONaraigh_2014}.

\subsection{Low density contrast}
\label{sec:nonlinear_low_density}
\begin{figure}[htbp]
	\centering
		\includegraphics[width=\textwidth]{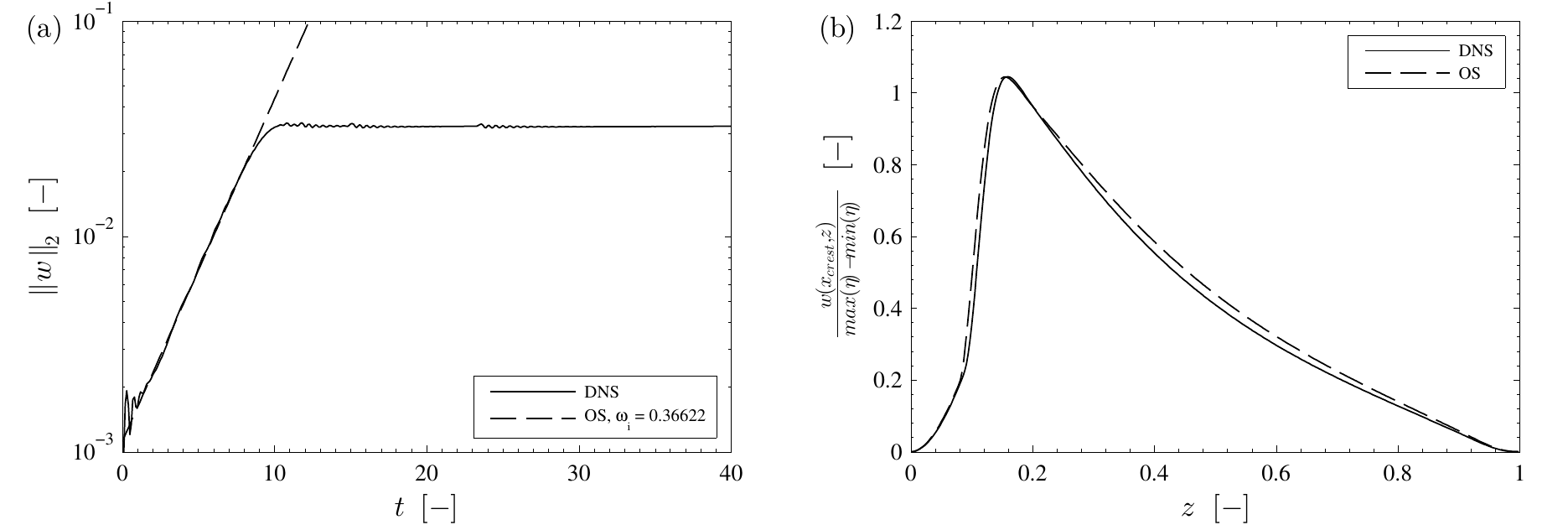}
	\caption{Comparison of nonlinear DNS with linear theory through semi-analytical Orr-Sommerfeld analysis for scenario dT4C. (a) shows the wave growth rate calculated as the $L^{2}$-norm of the perturbation $w$-velocity; (b) is a plot the streamfunction normalized by the wave height at the position of the wave crest at $t = 4.2$.}
	\label{fig:comparison_DNS_OS}
\end{figure}

To allow for rigorous comparison with linear theory, we perform direct numerical simulations for each flow regime of the low-density-contrast case (lower part of Table~\ref{tab:temporal_scenarios}). For that purpose, we set the wavenumber $\alpha_{n}$ in Eq.~\eqref{eq:initial_interface} equal to the linearly most unstable mode $\alpha_{m}^{temp}$ of the respective scenario. Figure~\ref{fig:comparison_DNS_OS}a shows the $L^{2}$-norm of the wall normal velocity perturbation $w$ over time for scenario dT4C and it can be seen that the disturbance grows with the rate $\omega_{i}^{temp}$ predicted by Orr-Sommerfeld theory.
Comparing the streamfunction $\phi_{j}\left(z\right)$ at the crest of the developing wave also yields excellent agreement (Fig.~\ref{fig:comparison_DNS_OS}b). Also the other scenarios follow the theoretical predictions in terms of growth rate in an equally convincing fashion (Table~\ref{tab:temporal_scenarios}).

\begin{figure}[htbp]
	\centering
		\includegraphics[width=\textwidth]{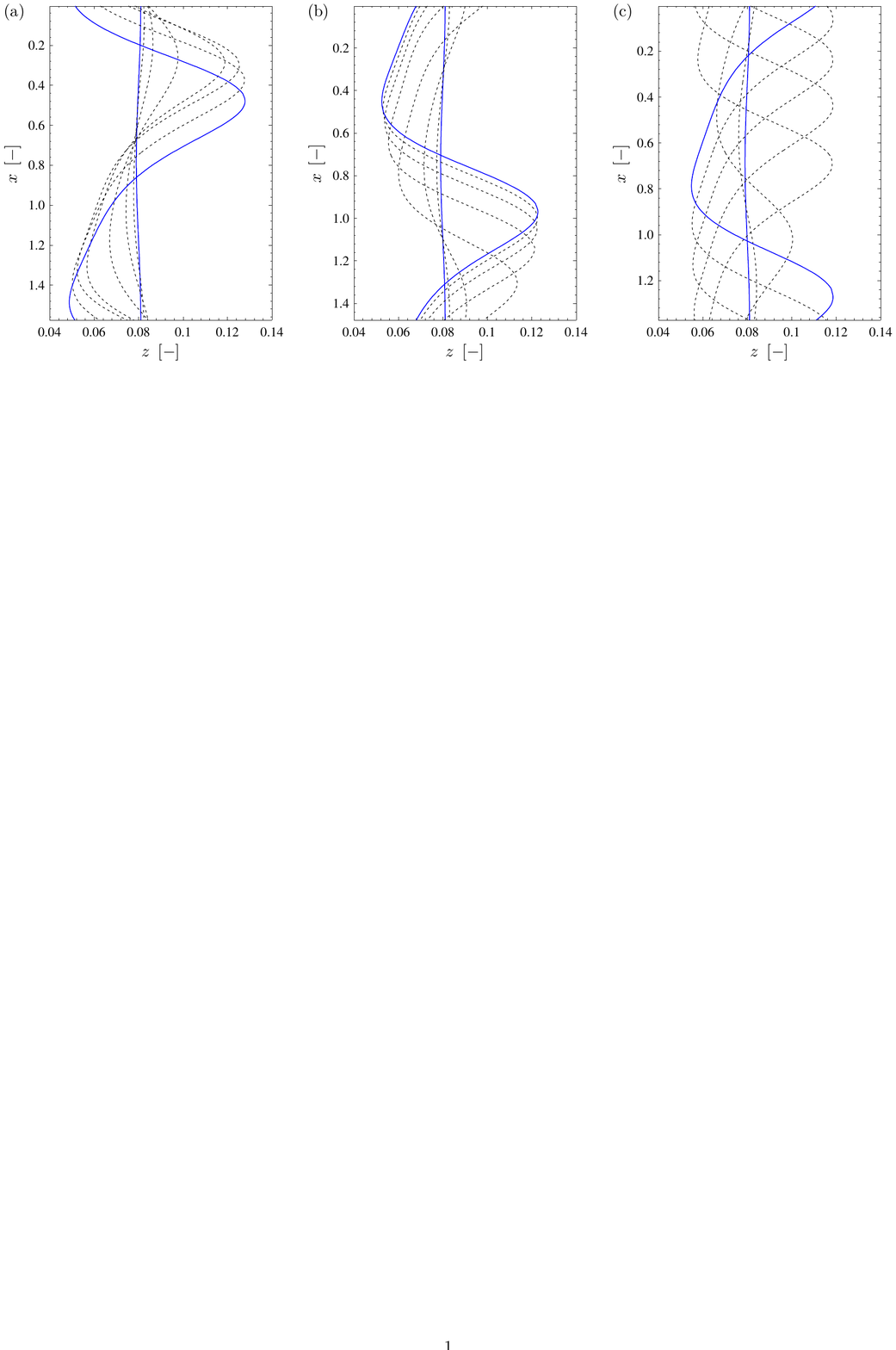}
	\caption{Typical evolution of the disturbed interface in the three flow regimes (low-density-contrast case) within the time interval $t = \left[0,17.5\right]$ at steps of $\Delta t = 2.5$ (dotted lines). The solid lines show the inital and final shape of the interface, respectively. (a) counter-current flow, dT4C; (b) loading, dT5L; (c) flooding, dT6F.}
	\label{fig:evolution_interfaces_low}
\end{figure}

\begin{figure}[htbp]
	\centering
		\includegraphics[width=\textwidth]{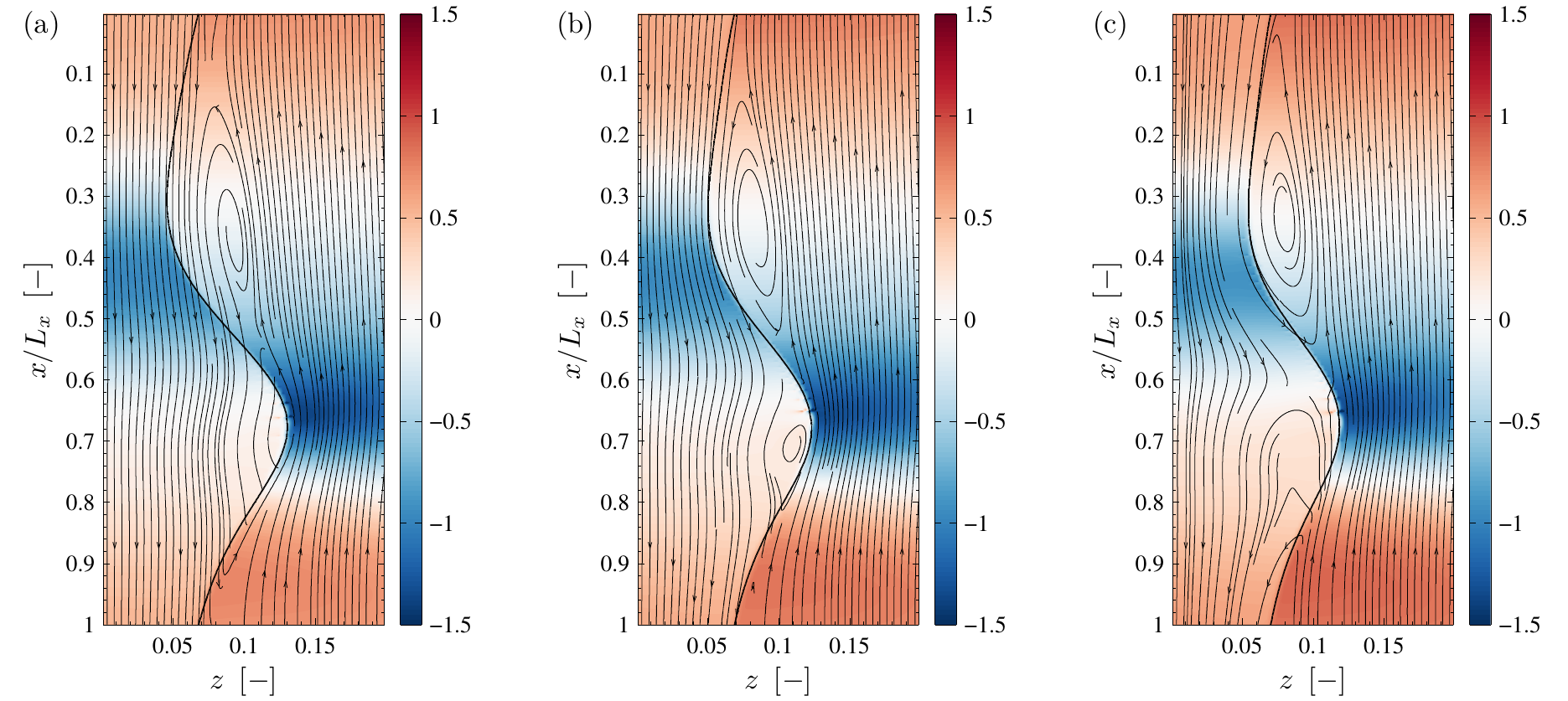}
	\caption{Pressure perturbation field, streamlines and liquid interface in a wall-fixed frame of reference for the scenarios near the loading curve (low-density-contrast case). (a) counter-current flow, dT4C, $t = 27.7$; (b) loading, dT5L, $t = 34.6$; (c) flooding, dT6F, $t = 23.3$.}
	\label{fig:comparison_scenarios_pressure_stream_inter_low}
\end{figure}

It can further be seen in Fig.~\ref{fig:comparison_DNS_OS}a that the wave initially grows exponentially before nonlinear effects gain importance at about $t = 8.0$. Beyond that point, wave growth slows down and eventually saturates, leading to a steady nonlinear wave of constant amplitude travelling along the interface. Figure~\ref{fig:evolution_interfaces_low}a depicts the early stage evolution of the interface up to saturation for the counter-current scenario.
It can be appreciated that the wave developing on the interface moves downwards as is predicted by linear theory. More insight into the flow features of the counter-current flow scenario is given by Fig.~\ref{fig:comparison_scenarios_pressure_stream_inter_low}a.
Plotted in a fixed frame of reference, a large anticlockwise rotating recirculation zone, positioned on the ``downwind'' side of the wave, is revealed in the gas phase. This vortex is sandwiched between the main flow of the gas and a thin layer of gas along the entire length of the interface, which is dragged downwards by the liquid due to interfacial shear stress. The liquid phase itself does not exhibit any major disturbances.

According to linear theory, the loading point of the system is reached by increasing the Froude number to $Fr = 1.179$ (scenario dT5L, see Fig.~\ref{fig:comparison_c_p_loading_curve_r1000_r10}b). At this point, the phase velocity of the wave vanishes as gravitational and lift forces on the wave hump balance each other, resulting in a standing wave. The evolution of this wave in its early stage is depicted in Fig.~\ref{fig:evolution_interfaces_low}b. It can be apprehended that the initial upward motion of the wave slows down after it is saturated and eventually comes to a complete halt around the position last indicated in the plot. As the system evolves further the saturated wave begins to move downwards with the bulk of the liquid (not shown here). This is attributed to nonlinear effects which lead to an anticipated deviation from the system behaviour predicted by linear theory. However, it is worth mentioning that the deviation is relatively small ($v_{p} = 0.0103$) and, hence, linear theory gives a good indication of the development and behaviour of the liquid interface for the loading scenario (dT5L). Moreover, the increased Froude number leads to a decrease of the mean liquid streamwise velocity, which, in turn, reduces the extent of the thin layer of downward gas flow along the liquid interface. Thus, the gas side vortex developing in the wave trough moves closer to the interface. The enhanced flow velocity on the gas side, on the other hand, stretches this eddy along the entire trough region (Fig.~\ref{fig:comparison_scenarios_pressure_stream_inter_low}b). On the liquid side, contrary to the counter-current scenario, a small anticlockwise rotating vortex occurs in the wave body, which indicates a negative (upwards) streamwise velocity of the liquid at the interface in the vicinity of the wave crest. However, the bulk of the liquid, which refers to the region of the film between channel wall and wave trough, remains largely unaffected.

By increasing the Froude number beyond the loading point, as in scenario dT6F, lifting forces on the wave body due to the enhanced gas flow overcome the gravitational forces, causing interfacial waves to travel upwards (Fig.~\ref{fig:evolution_interfaces_low}c). Similar to the previous scenarios, the developing wave saturates and forms a coherent structure. However, with higher Froude numbers the general trend of increasing growth rate and decreasing amplitude becomes apparent. Furthermore, the change of direction in which the interfacial wave propagates leads to a more agitated liquid film, especially in the wave body (Fig.~\ref{fig:comparison_scenarios_pressure_stream_inter_low}c). Compared to scenario dT5L, an extended region of liquid near the wave crest experiences upward movement with the wave body and becomes recirculated in an enlarged eddy. The bulk of the liquid, on the other hand, essentially remains unaltered also in this scenario.

\begin{figure}[htbp]
	\centering
		\includegraphics[width=\textwidth]{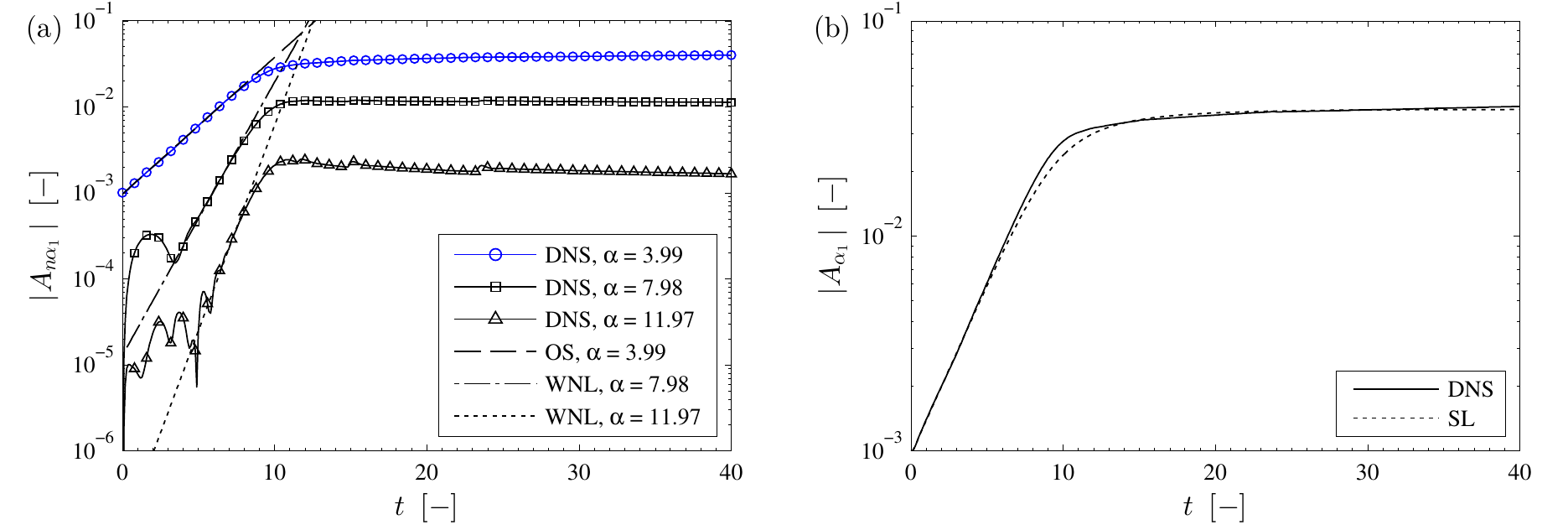}
	\caption{Comparison of direct numerical simulations with weakly nonlinear theory for the counter-current scenario (dT4C). (a) interfacial spectra; (b) Stuart-Landau equation, eq.~\eqref{eq:SL_equation}, fitted to fundamental mode.}
	\label{fig:weakly_nonlinear_counter}
\end{figure}

After having discussed the influence of the Froude number, i.e. gas flow rate, on the gas-liquid flow in a qualitative manner, we now want to turn our attention to a qualitative description of the nonlinear wave dynamics. To understand the genesis of these, we compute the spectra of the interface height for each time step. The spectra of the first three harmonics are shown in Fig.~\ref{fig:weakly_nonlinear_counter}a.
Matching the rate given by Orr-Sommerfeld theory, the first harmonic ($\alpha_{1} = \alpha_{m}^{temp} = 3.99$) grows exponentially fast at the beginning. At the same time, the higher harmonics ($\alpha_{2} = 7.98, \alpha_{3} = 11.97$) are linearly stable, as predicted by the same theory. However, these harmonics eventually also undergo exponential growth, whereat the $n^{th}$ harmonic grows at a rate $n\omega_{i}^{temp}\left(\alpha_{0}\right)$. As time goes by, the growth rate of all harmonics decreases and the amplitudes saturate simultaneously. It is thus apparent that the dynamics of the higher harmonics are ``slaved'' to the first harmonic. This temporal development of the interfacial spectra is perfectly consistent with weakly nonlinear theory~\cite{Barthelet_1995}. The framework of the same theory allows to model the temporal development of the first harmonic using the Stuart-Landau (SL) equation~\cite{Stuart_1958,Landau_1944}. Including up to quintic-order terms, the absolute value of the finite amplitude follows
\begin{equation}
	\frac{\mathrm{d}}{\mathrm{d}t}\left|A_{\alpha_{1}}\right|^{2}
																									= \mu_{1}\left|A_{\alpha_{1}}\right|^{2}+\mu_{2}\left|A_{\alpha_{1}}\right|^{4}+\mu_{3}\left|A_{\alpha_{1}}\right|^{6},
\label{eq:SL_equation}
\end{equation}
where $\mu_{1}$ is twice the temporal growth rate $2\omega_{i}^{temp}$ and $\mu_{2,3}$ is twice the real part of the Landau coefficients~\cite{Schmid_2001}. These coefficients have been fitted numerically for the different scenarios discussed above and are listed in Table~\ref{tab:landau_coeff} together with the root mean square deviation (RMSD) and R$^{2}$-value of the best fit, illustrating the excellent agreement between theory and direct numerical simulations.
\begin{table}[htbp]
	\begin{ruledtabular}
		\begin{tabular}{cccccc}
			Scenario & $\mu_{1}$ & $\mu_{2}$ & $\mu_{3}$ & RMSD & R$^{2}$\\
			\hline
			dT4C	& 0.7344	& -867.1		& 252355	& $1.1306\cdot 10^{-3}$	& 0.9932\\
			dT5L	& 0.9290	& -913.9		& 0							& $8.1751\cdot 10^{-4}$	& 0.9933\\
			dT6F	& 1.1582	& -1410.1	& 0							& $5.7185\cdot 10^{-4}$	& 0.9957\\
			dT7F	& 2.7613	& -5466.6	& 0							& $9.0486\cdot 10^{-4}$	& 0.9749\\
		\end{tabular}
	\end{ruledtabular}
	\caption{Landau coefficients (real part, fitted to DNS results) for the scenarios investigated under low density contrast.}
	\label{tab:landau_coeff}
\end{table}
The negative value of $\mu_{2}$ confirms the existence of the supercritical bifurcation observed in all scenarios. For scenario dT4C, the relatively large value of $\mu_{3}$ points towards a longer transition phase with decreasing growth of the wave amplitude leading from the initial stage of exponential growth to saturation in this scenario. In contrast, the wave in case dT5L, dT6F and dT7F reaches its saturated state faster, which is reflected by the vanishing of $\mu_{3}$ in Eq.~\eqref{eq:SL_equation}. The comparatively low R$^{2}$-value in scenario T7F can be explained by an overshoot of the amplitude in the transition phase due to transient effects before the wave settles at a stable equilibrium amplitude.

\subsection{High density contrast}
\label{sec:nonlinear_high_density}
\begin{figure}[htbp]
	\centering
		\includegraphics[width=0.95\textwidth]{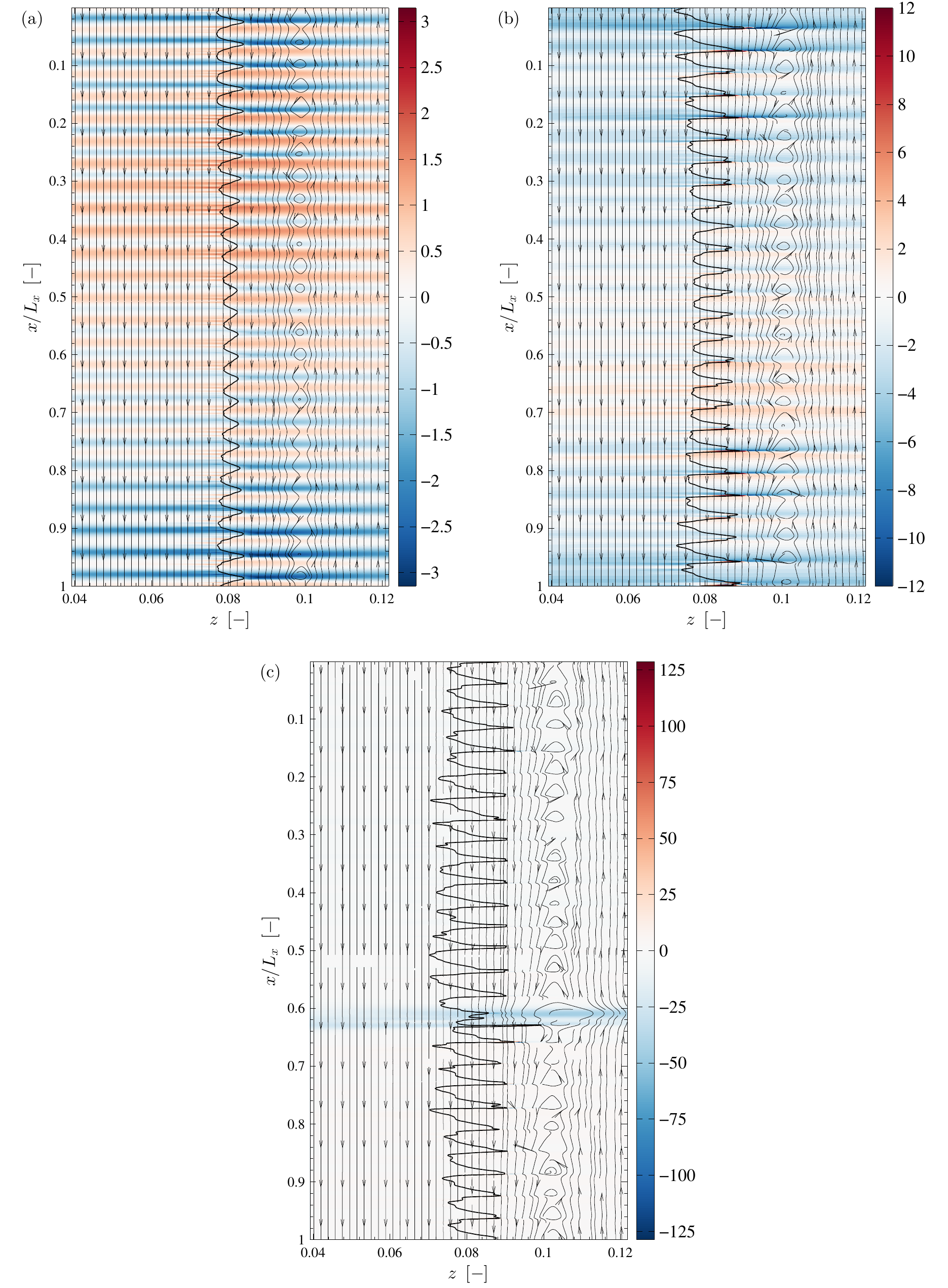}
	\caption{Pressure perturbation field, streamlines and liquid interface in a wall-fixed frame of reference for the scenarios near the loading curve (high density contrast). (a) $t = 1.08$; (b) $t = 1.71$; (c) $t = 2.025$.}
	\label{fig:comparison_scenarios_pressure_stream_inter_high}
\end{figure}

To gain insight in the nonlinear wave dynamics under high-density-contrast conditions, we perform direct numerical simulations with the system parameters corresponding to Fig.~\ref{fig:dispersion_relations_r1000}e, scenario T3C (Table~\ref{tab:temporal_scenarios}). This system exhibits two distinct linearly unstable modes, one long-wave shear mode and a short-wave interfacial mode, which are accounted for by $\alpha_{REY_{G}} = 1.549$ and $\alpha_{TAN} = 40.27$ in Eq.~\eqref{eq:initial_interface}, respectively. It is due to the short-wave nature of the interfacial mode combined with the high inertia of the liquid phase that the flow characteristics near the interface (Fig.~\ref{fig:comparison_scenarios_pressure_stream_inter_high}) differ significantly from those observed in the low-density-contrast case.
Unlike the scenarios presented in the previous section, no prominent features, such as recirculation zones, emerge within the wave train. However, a vortex layer with the familiar ``cat's eye structure'' develops at the demarcation between the bulk of the gas and a thin gas layer dragged downwards by the liquid due to interfacial shear stress. Thereby, the vortices are pinched between two consecutive high-pressure regions, which are forming on the downstream side of the short-wave crests. Further snapshots indicate that this vortex layer is unstable to secondary instability. Hence, the wave form in Figure~\ref{fig:comparison_scenarios_pressure_stream_inter_high} should not be regarded as a quasi-steady state.

\begin{figure}[htbp]
	\centering
		\includegraphics[width=\textwidth]{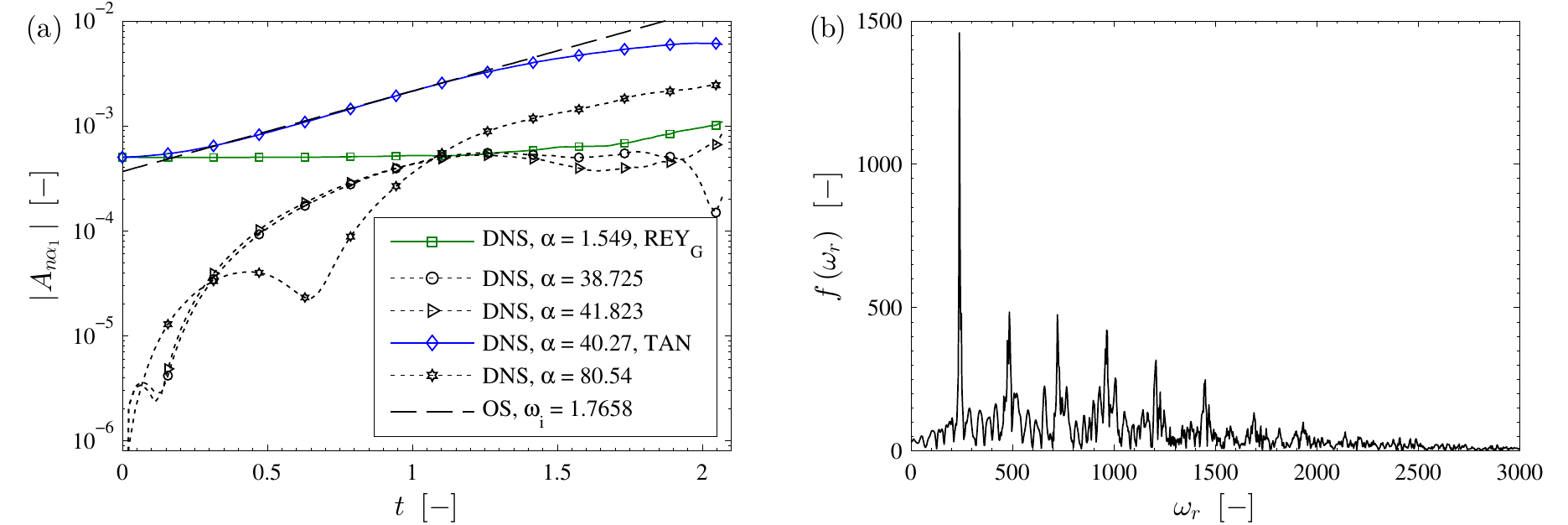}
	\caption{Spectral analyses confirming the relevance of the Stuart-Landau theory for the high-density-contrast case. (a) Spectra in the spatial domain showing the evolution of relevant wavenumber normal modes; (b) Power-spectral density $f(\omega)$ in the frequency domain.}
	\label{fig:spectral_analyses}
\end{figure}

Similar to the low-density-ratio case, the nonlinear dynamics of the interfacial mode appear to be consistent with Stuart-Landau theory, albeit these dynamics develop faster due to the higher growth rate of the first harmonic ($\alpha_{TAN} = 40.27$). Evidence is provided in Figure~\ref{fig:spectral_analyses}.
Consistency with the Stuart-Landau theory is especially clear in the time-frequency domain. In particular, in Figure~\ref{fig:spectral_analyses}b, a power-spectral density
\begin{equation}
	f\left(\omega_{r}\right) = 
		\left|\int_0^T w\left(x_0,\delta_L,t\right)\mathrm{e}^{-\mathrm{i}\omega_r t}\,\mathrm{d}t\right|,\qquad x_0 = 0.007
\label{eq:power_spectral_densisty}
\end{equation}
is shown ($T$ corresponds to the duration of the simulation). A well-defined global maximum is observed at at $\omega_{r} =239.8$, corresponding to the frequency of the most-dangerous TAN mode at $\alpha=40.27$ in the spatial domain. Successive maxima at $\omega_r=n(239.8)$ with $n=2,3,\cdots$ indicate that the ``overtones'' of the TAN mode are slaved to the TAN mode itself.

The maxima in the power-spectral density function, although well defined, are by no means sharp. The broadening of the maxima here is a sign that the wave train is not strictly periodic, and that a very large number frequencies is present in the dynamics. Crucially, the broadness of the lines is a function of the simulation time: for longer simulations, more and more frequencies come into play (albeit that the same maxima remain dominant throughout). This indicates the onset of chaos. Thus, the Stuart-Landau theory manifests itself as the leading-order approximation to a chaotic dynamics, albeit that the wave train in Figure~\ref{fig:comparison_scenarios_pressure_stream_inter_high} is subject to secondary instability.

Other second-order effects are key to understanding the secondary instability. The vortex layer in Fig.~\ref{fig:comparison_scenarios_pressure_stream_inter_high}a is not steady but breaks down at later times (Fig.~\ref{fig:comparison_scenarios_pressure_stream_inter_high}b). In particular, a long-wave perturbation to the vortex layer (wavelength on the domain scale) is seen to coincide with the significant growth of the long-wave shear mode. Therefore, it would appear that a complicated secondary instability sets in, involving a destabilization of the vortex layer by perturbations that are fed by the long-wavelength linearly unstable (but subdominant) mode. At this stage, individual waves steepen even further, up to the point where wave overturning is imminent (Fig.~\ref{fig:comparison_scenarios_pressure_stream_inter_high}b).

One may furthermore recall that the strong presence of a shear mode in the system is a signal of the supercritical transition to bulk turbulence in one of the phases. This is not directly relevant in the present scenario, as the secondary instability of the laminar wave train sets in before this transition. In either case, it is clear that secondary instability may inhibit the operation of the system in a quasi-steady laminar state at high density ratios.

\section{Conclusions}
\label{sec:conclusions}
\raggedbottom
We have presented a comprehensive study on two-dimensional laminar flow of a vertical liquid film sheared by laminar counter-current gas in a confined channel. This study tries to further elucidate the nature of interfacial instability in such two-phase flows using several complementary methods, namely Orr-Sommerfeld analysis, energy budget analysis as well as high resolution direct numerical simulations. Two sample systems have been selected for investigation: one with a high density contrast ($\rho_{L}/\rho_{G} = 1000$) and a second with a low density contrast ($\rho_{L}/\rho_{G} = 10$). In both cases, the same viscosity contrast ($\mu_{L}/\mu_{G} = 50$) and comparatively low surface tension ($\gamma$ = \SI{1e-3}{\newton\per\metre}) was used. In our study we focussed on analysing the influence of liquid film thickness and applied pressure drop on the development of interfacial waves.

Temporal linear stability analysis reveals that the liquid interface is inherently unstable for both cases. In the system with high density ratio short-wave instability is predominant, whereas the low-density-contrast case tends to favour long-wave instability. Furthermore, the instability is governed by a multitude of coexisting unstable modes (interfacial mode, shear mode in both phases, internal mode) under high-density-ratio conditions, where the latter two modes indicate the onset of turbulence in the bulk of one of the phases. In contrast, the low-density-ratio system exhibits only one unstable mode, which is consistent with the Yih mode~\cite{Yih_1967} and a manifestation of the viscosity contrast.

Additionally, we use the phase velocity of the linearly most unstable mode to identify two distinct flow patterns: counter-current flow and flooding. In the former regime, the phase velocity is positive, which relates to interfacial waves propagating downwards. By contrast, flooding is characterised by upward travelling waves, leading to partial flow reversal in the liquid film. During the transition from counter-current to flooding regime, standing waves evolve at the loading point. In the case with high density contrast, we found a region of flooding amidst the counter-current regime, which is the result of mode competition (shear mode in the gas layer vs. interfacial mode).

We have further determined the nature of the instability in the spatio-temporal framework under low-density-ratio conditions. Using this information, we established a second flow map indicating the transition from convective to absolute instability (C/A). Besides standard Orr-Sommerfeld analysis, we also adopted the analytical connection between temporal and spatio-temporal growth rates in the linear regime as presented in Reference~\cite{ONaraigh_2013}. This approach, which is based on analytical continuation, circumvents possible difficulties in identifying the absolute growth rate that are associated with the multivalued nature of the eigenvalue problem and specifics of the problem at hand. Compared with OS analysis and DNS, this method shows good agreement and is therefore appropriate to accurately estimate the absolute growth rate of the instability. We find that the system is absolutely unstable in most parts of the parameter space considered herein with the exception of two narrow bands at low applied pressure drop and low film thickness, respectively. In the high-density-ratio case, mode competition and mode coalescence between the multiple unstable modes hindered the mapping of convective and absolute instability in the respective parameter space. For such systems, linearized DNS~\cite{ONaraigh_2013a} may allow for more conclusive results.

To assess the development of interfacial waves up to finite amplitudes, we perform direct numerical simulations for parameters within the established flow regimes of the low-density-ratio case, using a level set method based solver that has been developed in-house. These simulations show excellent agreement with linear theory during the stage of exponential wave growth and confirm the determined flow patterns. The simulations also show saturation of the waves once nonlinear effects become important. A Fourier analysis reveals that the growth of the higher harmonics of the interfacial waves is coupled to that of the fundamental in a fashion which is consistent with weakly nonlinear theory. The growth of the first harmonic also agrees well with the Stuart-Landau model, thus underpinning the weakly nonlinear nature of the instability and, moreover, suggesting the existence of a supercritical bifurcation. Regarding high-density-contrast conditions, the dynamics of the (short-wave) interfacial mode appear to be similar to those observed under low density ratio. However, direct numerical simulations suggest that the emergence of an additional (long-wave) shear mode triggers a secondary instability, which leads to a chaotic wave train showing signs of imminent wave overturning. 

In summary, the combination of generic complementary (semi-)analytical and numerical methods presented herein yields a comprehensive and convincing characterization of interfacial instability in vertical counter-current gas-liquid flows. We are therefore confident that this rigorous methodology can be employed to further elucidate the dynamics of parallel shear flows in a wide range of technically relevant parameter regimes, such as flows with high viscosity contrast or non-steady gas flow. Beyond that, the outlined approach may be used as a starting point for the future study of heat and mass transfer phenomena in vertical gas-liquid flows.

\section*{Acknowledgements}
%
The authors gratefully acknowledge the financial support of Sulzer Chemtech Ltd as well as the Scottish Funding Council through the Scottish Energy Technology Partnership (Project ETP24). This work made use of the facilities of ARCHER, the latest UK National Supercomputing Service (\url{http://www.archer.ac.uk}). ARCHER provides a capability resource to allow researchers to run simulations and calculations that require large numbers of processing cores working in a tightly coupled, parallel fashion. The ARCHER Service is provided by the EPSRC, NERC, EPCC, Cray Inc. and the University of Edinburgh. Further access to the facilities of ARCHER was provided through the ARCHER Resource Allocation Panel, project number e174.

\appendix*
\section{Full formulation of the linear stability analysis and numerical solution}
\label{app:appendix}
\subsection{Base flow velocity profile}
\label{app:base_flow_velocity_profile}
Here, we give a detailed formulation of the governing equations underlying the linear stability analysis undertaken in \S\S~\ref{sec:temporal_stability} and~\ref{sec:A/C_instability} as well as a description of the numerical method used to solve the corresponding generalized complex eigenvalue problem.

Under the assumption of a steady, spatially uniform, laminar and incompressible flow in both phases, the Navier-Stokes equations governing the undisturbed base flow reduce to standard balances between pressure as well as viscous and gravitational forces. For the liquid film, this balance can be written as
\begin{equation}
	\mu_{L}\frac{\mathrm{d}^{2}\tilde{u}_{0}}{\mathrm{d}z^{2}}
	- \frac{\mathrm{d}p}{\mathrm{d}x}
	+ \rho_{L}g = 0,\quad	-d_{L}\leq\tilde{z}\leq 0,
\label{app:eq:liquid_NS}
\end{equation}
where $\tilde{u}_{0}$ is the dimensional base flow velocity in the respective phase (tildes denote dimensional quantities). Equation~\eqref{app:eq:liquid_NS} is subject to no-slip at the liquid side channel wall, $\tilde{z} = -d_{L}$, and continuity of tangential stress at the interface, $\tilde{z} = 0$:
\begin{equation}
	\tilde{u}_{0}\left(\tilde{z} = -d_{L}\right) = 0,\quad
	\left.\mu_{L}\frac{\mathrm{d}\tilde{u}_{0}}{\mathrm{d}\tilde{z}}\right|_{\tilde{z} = 0^{-}} = -\tau_{int}.
\label{app:eq:liquid_NS_BC}
\end{equation}
Thus, integration of Eq.~\eqref{app:eq:liquid_NS} yields
\begin{equation}
	\tilde{u}_{0}\left(\tilde{z}\right) =
	\frac{1}{2\mu_{L}}\left(\frac{\mathrm{d}p}{\mathrm{d}x}-\rho_{L}g\right)\left(\tilde{z}^{2}-d_{L}^{2}\right)
	- \frac{\tau_{int}}{\mu_{L}}\left(\tilde{z}+d_{L}\right),\quad -d_{L}\leq\tilde{z}\leq 0
\label{app:eq:liquid_velocity_profile_dim}
\end{equation}
as the velocity profile for the liquid film. The interfacial velocity, which constitutes one of the boundary conditions of the gas layer, reduces to
\begin{equation}
	\tilde{u}_{0,int} =
	\tilde{u}_{0}\left(\tilde{z} = 0\right) =
	-\frac{1}{2\mu_{L}}\left(\frac{\mathrm{d}p}{\mathrm{d}x}-\rho_{L}g\right)d_{L}^{2}
	- \frac{\tau_{int}}{\mu_{L}}d_{L}.
	\label{app:eq:base_flow_interfacial_velocity}
\end{equation}

The velocity profile for the laminar gas layer is derived analogous to the liquid layer. We therefore write the force balance as
\begin{equation}
	\mu_{G}\frac{\mathrm{d}^{2}\tilde{u}_{0}}{\mathrm{d}z^{2}}
	- \frac{\mathrm{d}p}{\mathrm{d}x}
	+ \rho_{G}g = 0,\quad	0 \leq\tilde{z}\leq d_{G},
\label{app:eq:gas_NS}
\end{equation}
which is subject to continuity of velocity and shear stress at the interface:
\begin{equation}
	\tilde{u}_{0}\left(\tilde{z} = 0\right) = \tilde{u}_{0,int},\quad
	\left.\mu_{G}\frac{\mathrm{d}\tilde{u}_{0}}{\mathrm{d}z}\right|_{\tilde{z} = 0^{+}} = -\tau_{int}.
\label{app:eq:gas_NS_IC}
\end{equation}
Applying these interfacial condition on Eq.~\eqref{app:eq:gas_NS} yields
\begin{equation}
	\tilde{u}_{0}\left(\tilde{z}\right) =
	\tilde{u}_{0,int} + \frac{1}{2\mu_{G}}\left(\frac{\mathrm{d}p}{\mathrm{d}x}-\rho_{G}g\right)\tilde{z}^{2}
	- \frac{\tau_{int}}{\mu_{G}}\tilde{z},\quad 0 \leq\tilde{z}\leq d_{G}
\label{app:eq:gas_velocity_profile_dim}
\end{equation}
as the gas side velocity profile. To determine the interfacial shear $\tau_{int}$ we apply the no-slip condition at the gas side channel wall, $\tilde{u}_{0}\left(\tilde{z} = d_{G}\right) = 0$:
\begin{equation}
	-\frac{1}{2\mu_{L}}\left(\frac{\mathrm{d}p}{\mathrm{d}x}-\rho_{L}g\right)d_{L}^{2}
	- \frac{\tau_{int}}{\mu_{L}}d_{L}\\
	+ \frac{1}{2\mu_{G}}\left(\frac{\mathrm{d}p}{\mathrm{d}x}-\rho_{G}g\right)d_{G}^{2}
	- \frac{\tau_{int}}{\mu_{G}}d_{G} = 0.
	\label{app:eq:root_shear_reynolds}
\end{equation}
In summary, the velocity profile of the undisturbed base flow in dimensional form reads
\begin{equation}
	\tilde{u}_{0}\left(\tilde{z}\right) =
	\left\{
	\begin{aligned}
		& \frac{1}{2\mu_{L}}\left(\frac{\mathrm{d}p}{\mathrm{d}x}-\rho_{L}g\right)\left(\tilde{z}^{2}-d_{L}^{2}\right)
						- \frac{\tau_{int}}{\mu_{L}}\left(\tilde{z}+d_{L}\right),&\quad-d_{L}\leq \tilde{z}\leq 0,	\\
		& -\frac{1}{2\mu_{L}}\left(\frac{\mathrm{d}p}{\mathrm{d}x}-\rho_{L}g\right)d_{L}^{2}
						- \frac{\tau_{int}}{\mu_{L}}d_{L}	\\
		& \phantom{\frac{1}{2\mu_{L}}\left(\frac{\mathrm{d}p}{\mathrm{d}x}-\rho_{L}g\right)\left(\tilde{z}^{2}-d_{L}^{2}\right)
												- \frac{\tau_{int}}{\mu_{L}}\left(\tilde{z}+d_{L}\right),}
				\mathllap{+ \frac{1}{2\mu_{G}}\left(\frac{\mathrm{d}p}{\mathrm{d}x}-\rho_{L}g\right)\tilde{z}^{2}
													- \frac{\tau_{int}}{\mu_{G}}\tilde{z},}&0 \leq \tilde{z}\leq d_{G}.
	\end{aligned}
	\right.
\label{eq:velocity_profile_dim}
\end{equation}
Applying the nondimensionalization scheme of Eq.\eqref{eq:dimensionless_variables} and \eqref{eq:dimensionless_parameters} finally leads to the dimensionless form of the base flow velocity profile as presented in Eq.~\eqref{eq:velocity_profile}.

\subsection{Perturbation equations}
\label{app:perturbation_equations}
As mentioned in \S~\ref{subsec:base_flow_LSA}, we introduce a small disturbance that shifts the flat interface from $z = 0$ to $z = \eta$, where $\left|\eta\right|\ll 1$. This (dimensionless) wave elevation gives rise to perturbations in the flow field of the form:
\begin{equation}
	\left(u, w, p\right) = \left(u_{0}\left(z\right)+\delta u\left(x, z, t\right), \delta w\left(x, z, t\right), p_{0}\left(z\right)+\delta p\left(x, z, t\right)\right),
\end{equation}
where the subscript zero denotes base flow quantities and the $\delta$ quantities are infinitesimally small perturbations. We use these variables of the flow field and obtain the linearized Navier-Stokes equations, which, after elimination of the pressure, further yield the linearized equation for the wall-normal vorticity $\delta\omega_{z}$ in both phases $\left(j = L, G\right)$:
\begin{equation}
	\left(\frac{\partial}{\partial t}+u_{0}\frac{\partial}{\partial t}-\frac{m_{j}}{r_{j}Re_{p}}\bm{\nabla}^{2}\right)\delta{\omega}_{z} = -\delta{w}\frac{\partial^{2}}{\partial z^{2}}u_{0},
\label{app:eq:vorticity_equation}
\end{equation}
with $\left(r_{L}, r_{G}\right)$ = $\left(r, 1\right)$, $\left(m_{L}, m_{G}\right)$ = $\left(m, 1\right)$ and
\begin{equation}
	\delta{\omega}_{y} = \frac{\partial}{\partial z}\delta{u} - \frac{\partial}{\partial x}\delta{w}.
\label{app:eq:normal_vorticity}
\end{equation}
It is further convenient to use the streamfunction representation $(\delta u,\delta w) = (\partial\Phi/\partial z,-\partial\Phi/\partial x)$ for the two-dimensional disturbance velocity field, hence
\begin{equation}
\delta\omega_z=\nabla^2\Phi.
\label{eq:app:vort}
\end{equation}%
 Assuming a wave-like solution of the form $\Phi(x,z,t) = \mathrm{e}^{\mathrm{i}(\alpha x-\omega t)}\phi(z)$ for the streamfunction, Equations~\eqref{app:eq:normal_vorticity} and~\eqref{eq:app:vort} lead to the Orr-Sommerfeld equations governing the stability of the liquid interface:
\begin{equation}
	\mathrm{i}\alpha
	\left[\left(u_{0}-\frac{\omega}{\alpha}\right)\left(\frac{\mathrm{d}^{2}}{\mathrm{d}z^{2}}-\alpha^{2}\right)\phi_{j}-\frac{\mathrm{d}^{2}u_{0}}{\mathrm{d}z^{2}}\phi_{j}\right] =
	\frac{m_{j}}{r_{j}Re_{p}}\left(\frac{\mathrm{d}^{2}}{\mathrm{d}z^{2}}-\alpha^{2}\right)^{2}\phi_{j},
\label{app:eq:orr_sommerfeld}
\end{equation}
where $\alpha = \alpha_{r} + \mathrm{i}\alpha_{i}$ and $\omega = \omega_{r} + \mathrm{i}\omega_{i}$ are the complex wavenumber and angular frequency, respectively. This equation is subject to no-penetration and no-slip conditions at both channel walls:
\begin{equation}
	\phi\left(-\delta_{L}\right) =
	\frac{\mathrm{d}}{\mathrm{d}z}\phi\left(-\delta_{L}\right) =
	\phi\left(\delta_{G}\right) =
	\frac{\mathrm{d}}{\mathrm{d}z}\phi\left(\delta_{G}\right) = 0.
\label{app:eq:BC_OS}
\end{equation}
Furthermore, conditions for continuity of velocity and tangential stress as well as a jump in the normal stress due to surface tension are applied to match the streamfunction across the interface at $z = 0$ (we use the notation $c = \omega/\alpha$):
\begin{subequations}\label{app:eq:IC_OS}
	\begin{equation}
		\phi_{L} = \phi_{G},
		\label{app:eq:IC1}
	\end{equation}
	\begin{equation}
		\frac{\mathrm{d}\phi_{L}}{\mathrm{d}z} =
		\frac{\mathrm{d}\phi_{G}}{\mathrm{d}z} +
																											\frac{\phi_{L}}{c-u_{0}}\left(\left.\frac{\mathrm{d}u_{0}}{\mathrm{d}z}\right|_{0^{+}}-
																											\left.\frac{\mathrm{d}u_{0}}{\mathrm{d}z}\right|_{0^{-}}\right),
		\label{app:eq:IC2}
	\end{equation}
	\begin{equation}
		m\left(\frac{\mathrm{d}^{2}}{\mathrm{d}z^{2}}+\alpha^{2}\right)\phi_{L} =
		\left(\frac{\mathrm{d}^{2}}{\mathrm{d}z^{2}}+\alpha^{2}\right)\phi_{G}
																	+\frac{\phi_{L}}{c-u_{0}}\left(\left.\frac{\mathrm{d}^{2}u_{0}}{\mathrm{d}z^{2}}\right|_{0^{+}}-
																	\left.\frac{\mathrm{d}^{2}u_{0}}{\mathrm{d}z^{2}}\right|_{0^{-}}\right),
		\label{app:eq:IC3}
	\end{equation}
	\begin{align}
		m\left(\frac{\mathrm{d}^{3}\phi_{L}}{\mathrm{d}z^{3}}
								-3\alpha^{2}\frac{\mathrm{d}\phi_{L}}{\mathrm{d}z}\right)
		+\mathrm{i}\alpha rRe_{p}\left(c-u_{0}\right)\frac{\mathrm{d}\phi_{L}}{\mathrm{d}z}
		+\mathrm{i}\alpha rRe_{p}\left.\frac{\mathrm{d}u_{0}}{\mathrm{d}z}\right|_{0^{-}}\phi_{L}
		-\frac{\mathrm{i}\alpha Re_{p}}{c-u_{0}}\frac{\alpha^{2}}{We}\phi_{L}\nonumber\\
		=\left(\frac{\mathrm{d}^{3}\phi_{G}}{\mathrm{d}z^{3}}
								-3\alpha^{2}\frac{\mathrm{d}\phi_{G}}{\mathrm{d}z}\right)
		+\mathrm{i}\alpha Re_{p}\left(c-u_{0}\right)\frac{\mathrm{d}\phi_{G}}{\mathrm{d}z}
		+\mathrm{i}\alpha Re_{p}\left.\frac{\mathrm{d}u_{0}}{\mathrm{d}z}\right|_{0^{+}}\phi_{G}.
		\label{app:eq:IC4}
	\end{align}
\end{subequations}
Using operator notation to rewrite Eq.~\eqref{app:eq:orr_sommerfeld}-\eqref{app:eq:IC_OS} yields
\begin{equation}
	\mathcal{L}\phi = \mathrm{i}\omega\mathcal{M}\phi,
\label{app:eq:eigenvalue_operator}
\end{equation}
which highlights the generalized eigenvalue problem associated with the stability problem.

\subsection{Numerical method}
\label{app:numerical_method}
We solve this eigenvalue equation numerically by employing a standard Chebyshev collocation method~\cite{Boomkamp_1997} with an approximation for the streamfunction of the form
\begin{equation}
	\phi\left(z\right)\approx
	\left\{
	\begin{aligned}
		& \sum_{n=0}^{N_{L}} a_{n}T_{n}\left(\frac{2z}{\delta_{L}}+1\right)\text{,\enspace}-\delta_{L}\leq z\leq 0,\\
		& \sum_{n=0}^{N_{G}} b_{n}T_{n}\left(\frac{2z}{\delta_{G}}-1\right)\text{,\enspace}\phantom{-\delta_{L}}0\leq z\leq \delta_{G},
	\end{aligned}
	\right.
\label{app:eq:trial_solution}
\end{equation}
where $T_{n}\left(\cdot\right)$ is the $n^{\mathrm{th}}$ Chebyshev polynomial of the first kind. We substitute this trial solution into Eq.~\eqref{app:eq:eigenvalue_operator} and evaluate the result at $N_{L}+N_{G}-6$ Gauss-Lobatto collocation points. Together with the eight boundary and interfacial conditions, this yields $N_{L}+N_{G}+2$ linear equations in as many unknowns. In matrix terms, we solve
\begin{equation}
	\bm{L}\bm{v} = \mathrm{i}\omega\bm{M}\bm{v},
\label{app:eq:eigenvalue_matrix}
\end{equation}
where $\bm{v} = \left(a_{0},\dots,a_{N_{L}},b_{0},\dots,b_{N_{G}}\right)^{\mathrm{T}}$. We use a standard linear algebra package (MATLAB\textsuperscript{\textregistered}) to solve this equation, thereby adjusting the number of collocation points $\left(N_{L}+1,N_{G}+1\right)$ until convergence is reached.


\end{document}